\documentclass[11pt,a4paper]{article}

\usepackage{hyperref}
\usepackage[latin1]{inputenc}
\usepackage{amsmath} 
\usepackage{amsfonts} 
\usepackage[dvips]{graphicx} 
\usepackage{url} 
\usepackage{color} 
\usepackage{mathrsfs} 
\usepackage{rotating}
\usepackage{amssymb}

\newcommand{\beq}{\begin{equation}}
\newcommand{\eeq}{\end{equation}}
\newcommand{\bea}{\begin{eqnarray}}
\newcommand{\eea}{\end{eqnarray}}
\newcommand{\ba}{\begin{array}}
\newcommand{\ea}{\end{array}}
\newcommand{\bec}{\begin{center}}
\newcommand{\eec}{\end{center}}
\newcommand{\bei}{\begin{itemize}}
\newcommand{\eei}{\end{itemize}}

\newcommand{\eq}[1]{Eq.~(\ref{#1})}
\newcommand{\Ref}[1]{Ref.~\cite{#1}}

\begin{document}

\begin{flushright}
\small
L2C:13-045 \\ LUPM:13-004
\end{flushright}

\vspace{0.3cm}
\begin{center}
{\LARGE \bf Minimal lepton flavour structures}\\
\vspace{0.2cm}
{\LARGE \bf lead to non-maximal 2-3 mixing}\\

\vspace{.8 cm}

{\large Michele Frigerio\,$^{a}$ and  Albert Villanova del Moral\,$^{b}$}

\vspace{.5cm}

\centerline{$^a$ {\it Laboratoire Charles Coulomb, UMR 5221 (CNRS/Universit\'e Montpellier 2)}}
\centerline{\it F-34095 Montpellier, FRANCE}
\vspace{0.2cm}
\centerline{$^b${\it Laboratoire Univers et Particules de Montpellier, UMR 5299 (CNRS/Universit\'e Montpellier 2)}}
\centerline {\it F-34095 Montpellier, FRANCE}

\end{center}
\vspace{.2cm}

\begin{abstract}
 
\noindent Present data prefer a large but non-maximal $2-3$ mixing in the lepton sector.
We argue that this value, in connection with $\sin\theta_{13}\simeq 0.15$, is the generic outcome of minimal flavour structures. 
We present a few different incarnations of this statement, in terms of lepton mass matrices depending on a small number of parameters, that
can be justified by discrete flavour symmetries.
We also propose a general procedure to study the correlation between $\theta_{23}$, the absolute scale and ordering of the neutrino masses,
and the leptonic CP-violating phases.

\end{abstract}

\section{Introduction}

Flavour models aim to explain the observed pattern of fermion masses and mixing, in terms of some dynamical mechanism for flavour symmetry breaking.
In the quark sector the regular hierarchy of masses and mixing angles definitely points to an underlying dynamics, that unfortunately may be difficult to probe directly.   
In the lepton sector, in the last 15 years neutrino oscillation experiments led  to a tremendous improvement in the knowledge of the neutrino
masses and mixing.
Still a few important observables in the neutrino mass matrix are presently unmeasured, and one expects substantial improvements in the next decade.

A reasonable (but not exclusive) figure of merit for flavour models is the ability to predict a set of observables in terms of a smaller set of parameters.
Speculations on ``symmetric" values of the lepton observables (maximal or zero mixing angles, vanishing or degenerate mass eigenvalues, \dots) 
have been pursued,
and recent indications against these extreme values may seem to disfavour most theoretical proposals. However, one should keep in mind that minimal flavour structures
do {\it not} imply extreme values of the observables, as we will show in a few explicit examples.

Oscillation experiments are sensitive, in particular, to the three lepton mixing angles. Before 2012, only $\theta_{12}$ and $\theta_{23}$
were known to be non-zero and large. One year later, the situation changed dramatically: the $1-3$ mixing angle $\theta_{13}$ 
has been precisely measured by reactor experiments
\cite{An:2012eh} to be of the order of the Cabibbo angle and, 
given the precise determination of $\theta_{12}$ by solar experiments, the largest uncertainty pertains to $\theta_{23}$,
that could significantly deviate from the maximal value. Indeed, global fits \cite{Tortola:2012te,GonzalezGarcia:2012sz} 
favour non-maximal $2-3$ mixing at the $2\sigma$ level (an effect driven by accelerator
neutrino data), with a slight preference for $\theta_{23}<\pi/4$ in the case of normal ordering of the mass spectrum (as a consequence of atmospheric neutrino data).
Before the measurement of a non-zero $\theta_{13}$,  
a considerable theoretical effort was invested in explaining a
 tri-bi-maximal pattern for the lepton mixing angles (see Ref.~\cite{Altarelli:2010gt} for a review).
In the latter approach, one needs now to correct significantly the leading-order predictions, by taking into account a number of sub-leading effects.
After the measurement of a relatively large $\theta_{13}$, several flavour models have been proposed, that predict a non-zero value of $\theta_{13}$
at leading order \cite{Ma:2011yi,Hernandez:2012ra}.

In this paper we would like to reconsider the lepton flavour structure in the light of the up-to-date knowledge of the mixing angles. 
Our focus will be on the non-maximal value of $\theta_{23}$ and on the search for the minimal flavour patterns that lead to such value.
Rather than elaborating on general aspects of model-building with flavour symmetries, we will modestly propose a few simple scenarios that make
some definite predictions.
In fact, as the number of technical assumptions in the models is reduced, we will find that 
the values of $\theta_{12}$ and $\theta_{13}$ point to a deviation from maximal $\theta_{23}$ of the required size.

The predictions of flavour models are meant to motivate the future experimental program, including precise neutrino oscillation experiments, as well
as measurements of the absolute neutrino mass scale and of the neutrino-less double-beta ($0\nu2\beta$) decay rates. In this spirit, we will consider lepton mass matrices that
depend on a small number of parameters, and we will propose a systematic procedure to illustrate the associated predictions (section 2).
We focus our attention on the mass matrix structures where the $2-3$ mixing angle is predicted close to the present best fit value, $\sin^2\theta_{23}\simeq 0.4$ or 
$\sin^2\theta_{23}\simeq 0.6$.
We will demonstrate that they can be associated to simple, spontaneously broken, discrete flavour symmetries.
One scenario corresponds to neutrino mass matrices with two zero entries and two non-zero entries equal to each other (section 3).
A second scenario resorts to the sum of two contributions to the neutrino mass matrix: a flavour off-diagonal term, plus a flavour-universal one (section 4). 
Finally, the minimal model with a flavour symmetry broken by a flavon doublet is presented (section 5).
These models allow, in particular, to correlate the $2-3$ mixing octant with the value of the CP violating phase $\delta$, as well as with  the ordering
of the neutrino mass spectrum (normal or inverse), thus motivating the experimentalists to solve the degeneracies 
that affect the long-baseline oscillation program \cite{Barger:2001yr}.

\section{From the neutrino mass matrix to the \\ observables
\label{preds}}

We assume the Standard Model is extended to include a Majorana mass matrix for the three active neutrinos.
The nine flavour observables in the neutrino sector can be currently divided in two groups.
On the one hand, four physical parameters are precisely determined (with less than 10\% uncertainty):
\beq
p_a= \Delta m^2_{21},~|\Delta m^2_{31}|,~\theta_{12},~\theta_{13}, 
\label{pa}\eeq
with the sign of $\Delta m^2_{31}$ still unknown.
On the other hand, five parameters are at most weakly constrained. The mixing angle $\theta_{23}$ can largely depart from the maximal value, $0.34\lesssim 
\sin^2\theta_{23} \lesssim 0.67$ at $3\sigma$ \cite{Tortola:2012te,GonzalezGarcia:2012sz}.
The lightest neutrino mass $m_{\text{light}}$ can
vary between zero and a few tenths of an electronvolt, the upper bound depending on the combination of 
cosmological data; we will assume a conservative limit, $\sum_i m_i \lesssim 0.5$ eV at 95\% C.L. \cite{Lesgourgues:2012uu}.
The Dirac-type CP-violating phase $\delta$ is unknown and can lie anywhere in the interval $[0,2\pi)$, with some range of values already disfavoured at the $1\sigma$ level
by global fits of oscillation data \cite{Tortola:2012te,GonzalezGarcia:2012sz}.
Finally, the two Majorana-type CP-violating phases are unknown as well.
The only experimentally accessible quantity sensitive to the latter two phases is the effective $0\nu2\beta$-decay mass parameter $m_{ee}$,
which is experimentally known to be smaller than a few tenths of an electronvolt; we will adopt the upper bound $m_{ee}\lesssim 0.38$ eV at 90\% C.L., 
coming from the EXO-200 experiment \cite{Auger:2012ar}.
The central objective of the present and future experimental program is the precise determination of these four observables (together with the sign of $\Delta m^2_{31}$):
\beq
x_i = \theta_{23},~ m_{\text{light}},~ \delta,~ m_{ee}~.
\label{xi}\eeq

Flavour symmetry models may predict special structures of the neutrino mass matrix $M_\nu$
(in the basis where the charged lepton mass matrix $M_e$ is diagonal), that depend on few parameters,
and therefore predict correlations between the observables.
In particular, four real parameters of $M_\nu$ can be fixed to reproduce the observed values of the $p_a$'s,
up to their small experimental uncertainties.
If $M_\nu$ depends on four (or less) parameters only, then one can derive a unique prediction for all the other observables, 
$x_i^{\pm} = x_i^{\pm}(p_a)$,
where the superscript denotes the sign of $\Delta m^2_{31}$. In this case the precise measurement of one $x_i$ can rule out the given matrix structure.
If $M_\nu$ depends on five parameters, then it is sufficient to specify the value of an additional observable $x_j$ to predict the values of the other three,
$x^{\pm}_i = x_i^{\pm}(x_j,p_a)$, $i\ne j$. Such matrix structures can be tested in general by measuring with sufficient precision two 
$x_i$'s. And so on and so forth.

If the procedure to compute $x_i^\pm$ is carried out analytically, then it provides the full set of predictions for a given form of $M_\nu$. 
In general such procedure requires some tedious algebraic manipulations, but the advantage is to avoid
a numerical sampling of the allowed range for the input parameters, or an expansion in powers of some small parameter, with the
associated ambiguities.

Let us illustrate our method for a specific matrix structure that will be useful in the following, and that depends on five physical parameters:
\begin{equation}
M_\nu=\left(\begin{array}{ccc} 
a & b & c \\
b & 0 & d \\
c & d & 0
\end{array}\right)~.
\label{C}
\end{equation}
The entries $a,\ b,\ c$ and $d$ are in general complex, but three phases can be absorbed in a redefinition of the three neutrino wavefunctions. 
Thus, there is a one-to-one correlation between the sought-after observables $x_i$'s. 
The six functions $x_i^{\pm} = x_i^{\pm}(x_j)$ for $i\ne j$ are shown in Figures 1 and 2, for the case `$+$'  ($m_{\text{light}}=m_1$, normal ordering) 
and `$-$' ($m_{\text{light}}=m_3$, inverse ordering), respectively.
The procedure to derive $x_i^\pm(x_j)$ analytically is described in the Appendix, 
and it applies to any matrix $M_\nu$ with two zero entries.
In principle one may find an analogue procedure for any structure of $M_\nu$ that depends on five parameters
(see  section \ref{sec:zeeplusd} for a particularly interesting example).

\begin{figure}[tbp]
\begin{tabular}{rr}
\includegraphics[width=0.28\textheight]{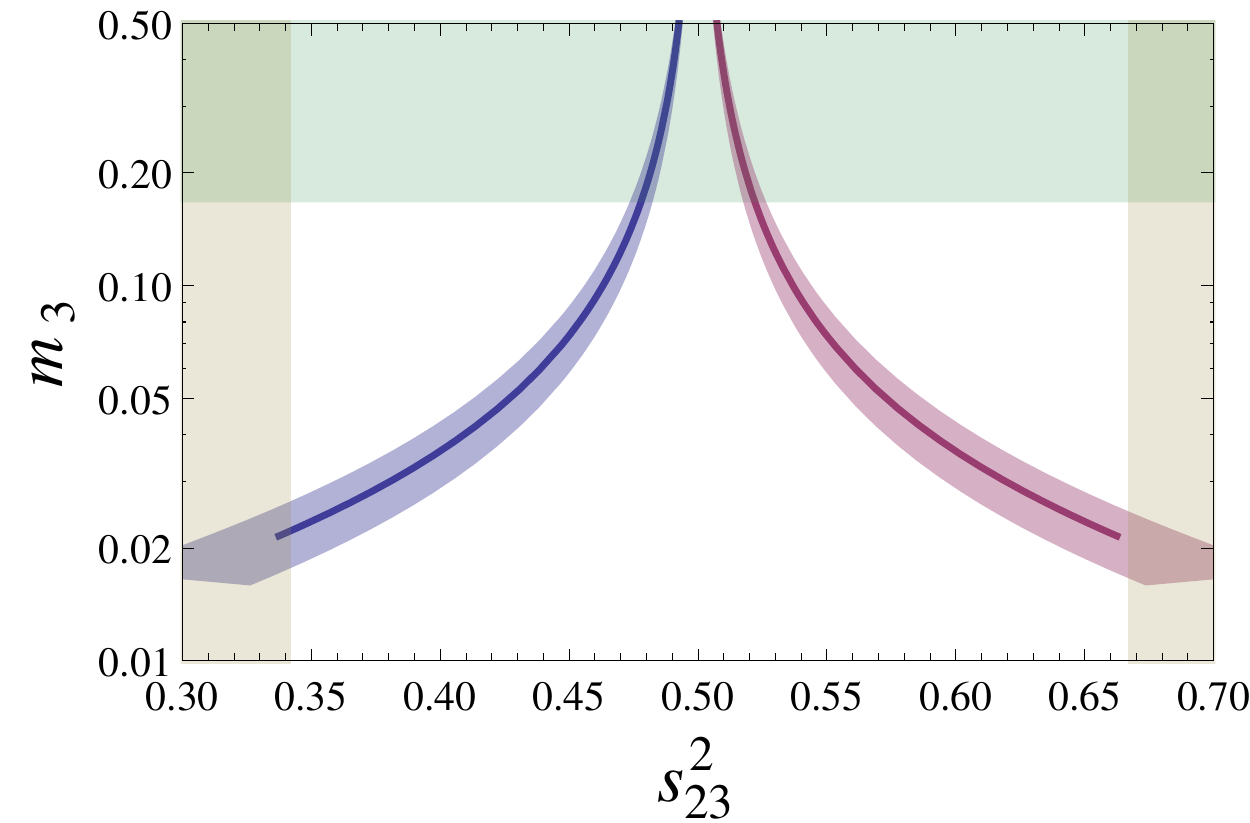}&
\includegraphics[width=0.28\textheight]{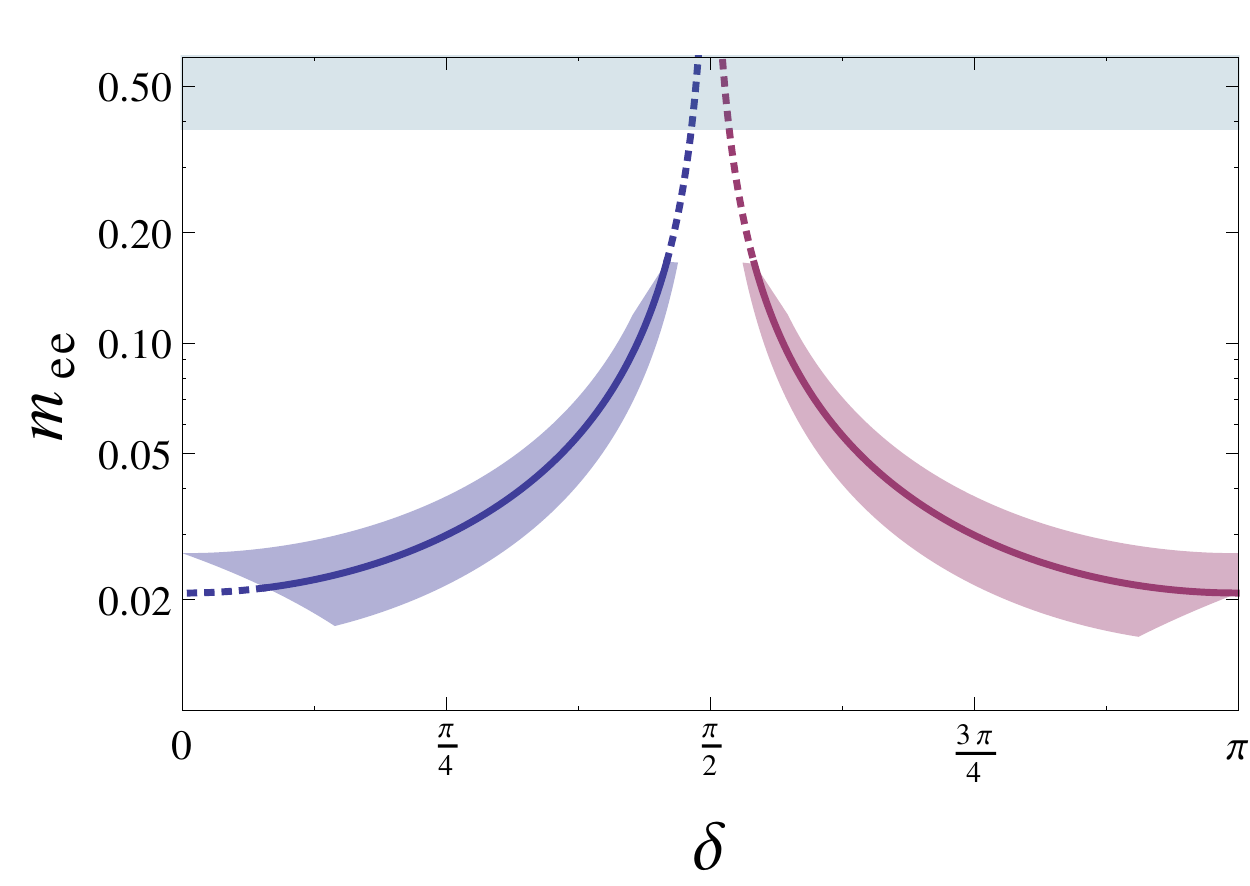}\\
\includegraphics[width=0.28\textheight]{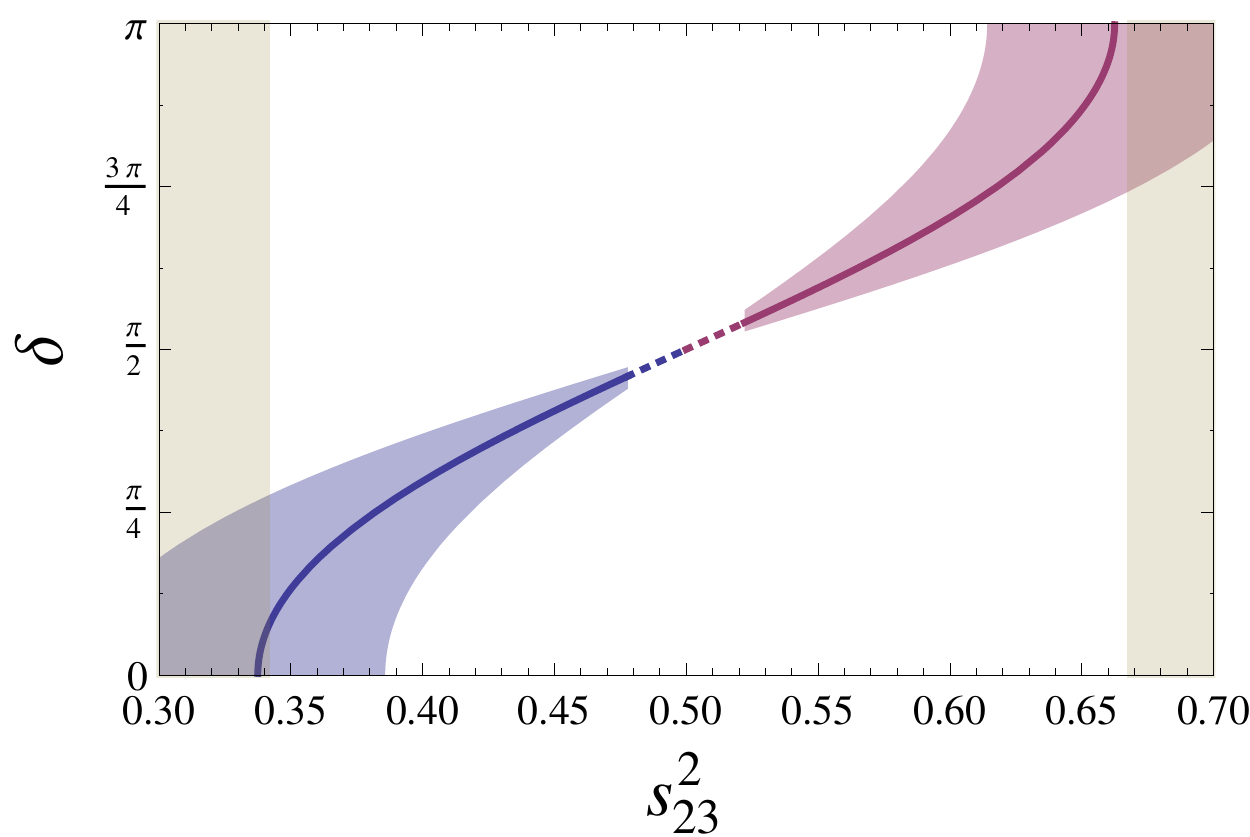} &
\includegraphics[width=0.28\textheight]{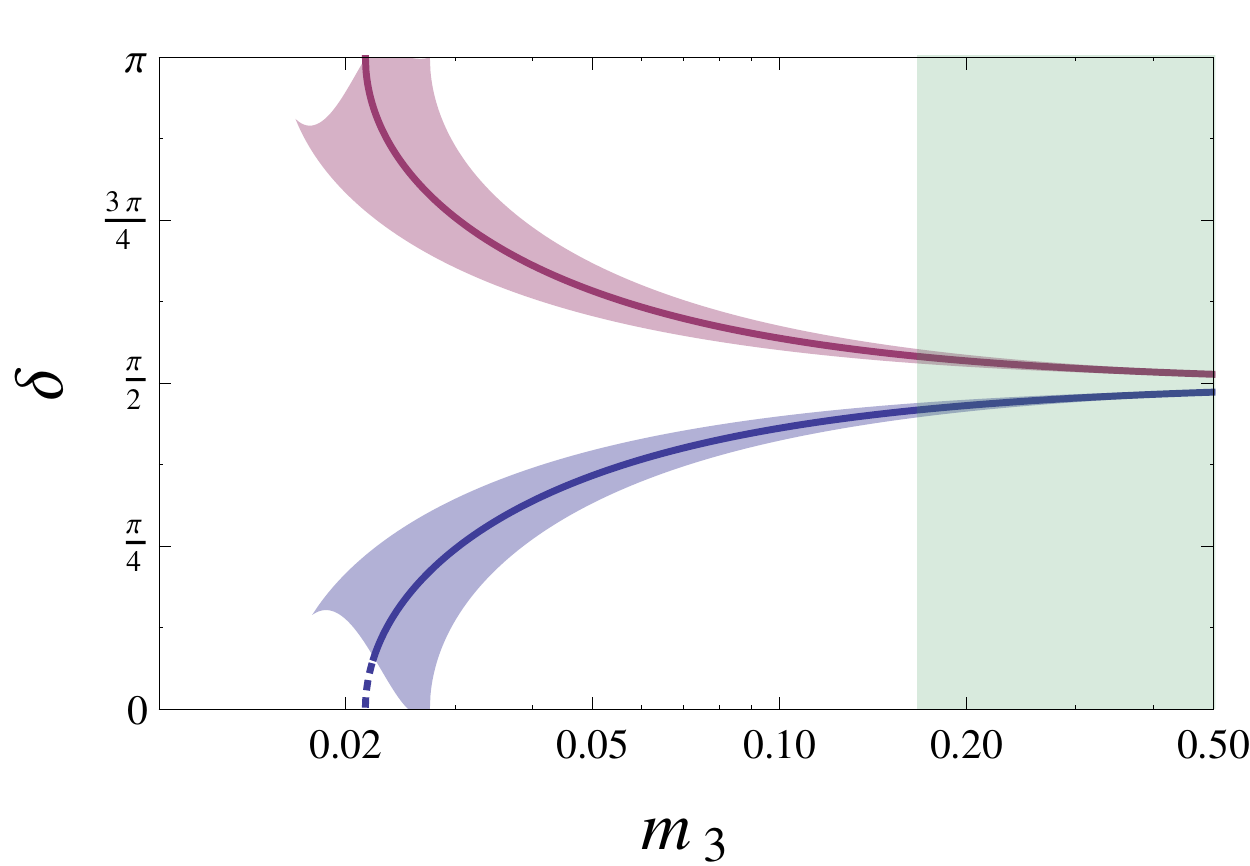}\\
\includegraphics[width=0.28\textheight]{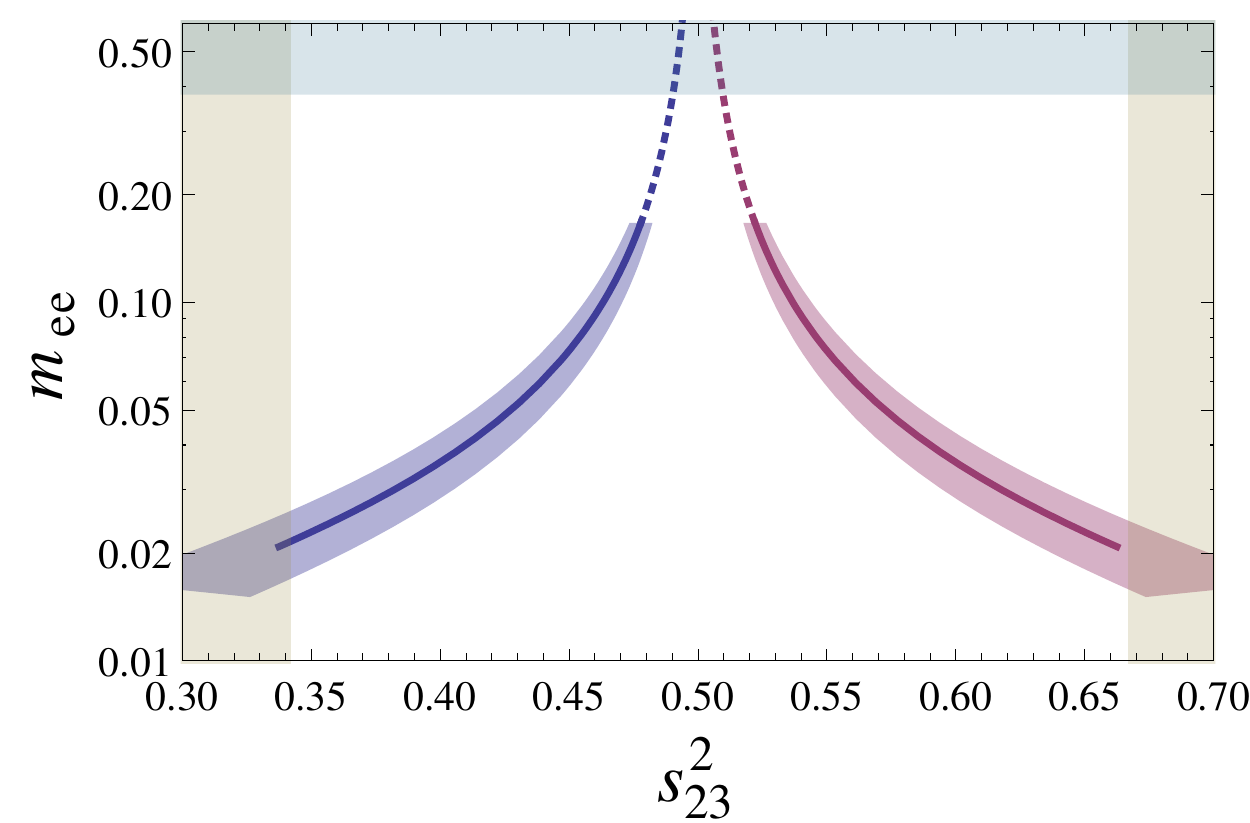}&
\includegraphics[width=0.28\textheight]{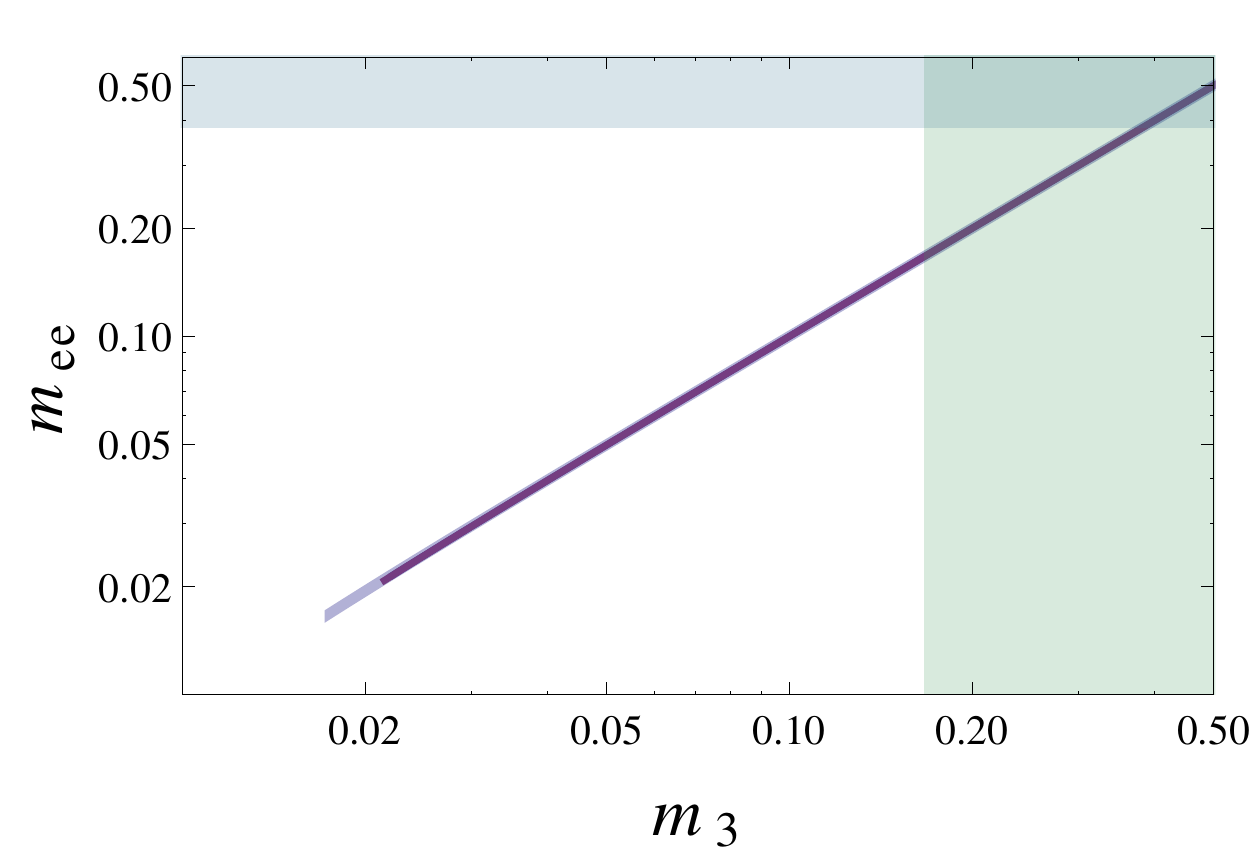}
\end{tabular}
\caption{\it
The correlations between $\sin^2\theta_{23}$, $\delta$, $m_3$ (in eV) and $m_{ee}$ (in eV) for the neutrino mass matrix
in \eq{C}, in the case of inverted ordering of the mass spectrum.
The thick purple lines correspond to the best fit value of the parameters $p_a$'s in \eq{pa}, 
while the purple shaded regions correspond to the $3\sigma$ allowed range for the $p_a$'s. 
The yellow bands are excluded by oscillation experiments.
The green (blue) bands are excluded by the cosmological upper bound on $\sum m_i$
(by the EXO-200 upper bound on $m_{ee}$). 
These three exclusion bands are taken into account in all the six panels:
the excluded portions of the best fit lines are dashed
and the excluded portions of the $3\sigma$ regions are not shaded. 
}\label{Cinverse}\end{figure}

\begin{figure}[tb]
\begin{tabular}{rr}
\includegraphics[width=0.28\textheight]{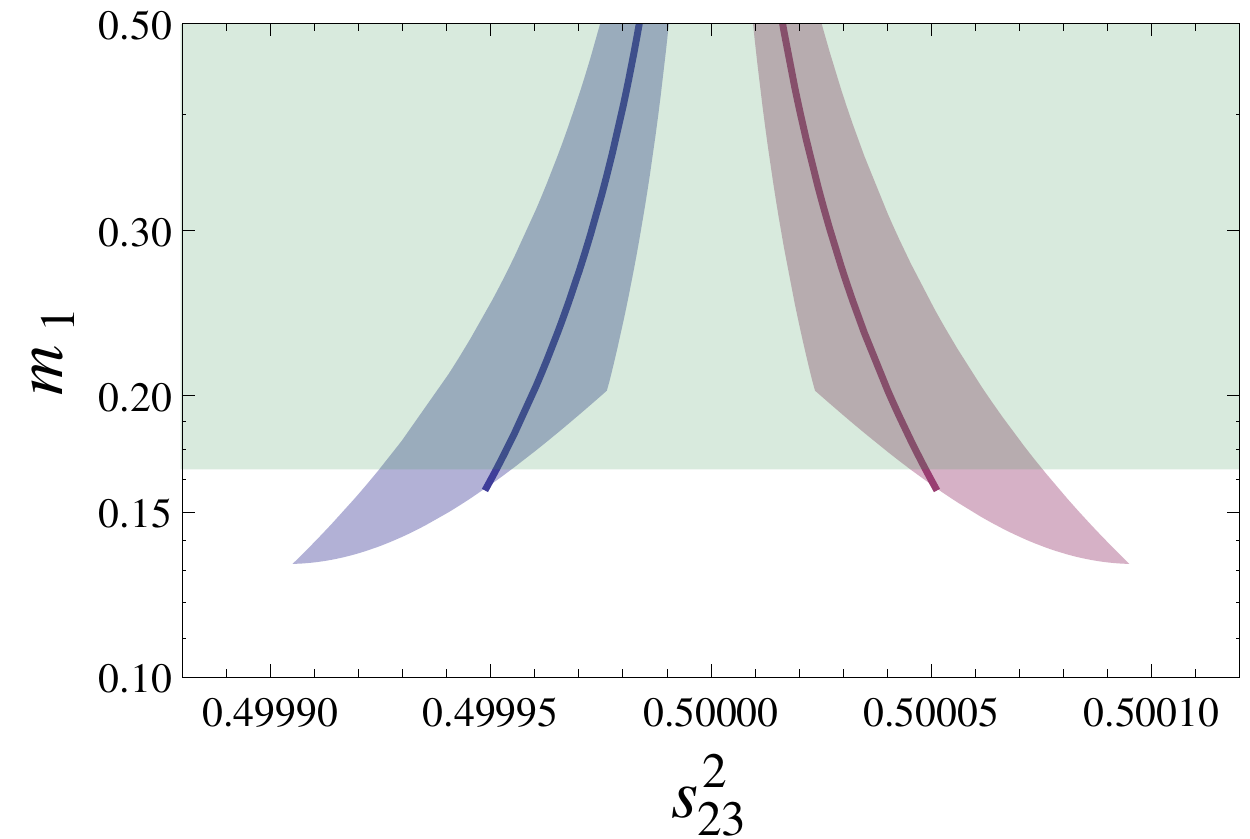}&
\includegraphics[width=0.28\textheight]{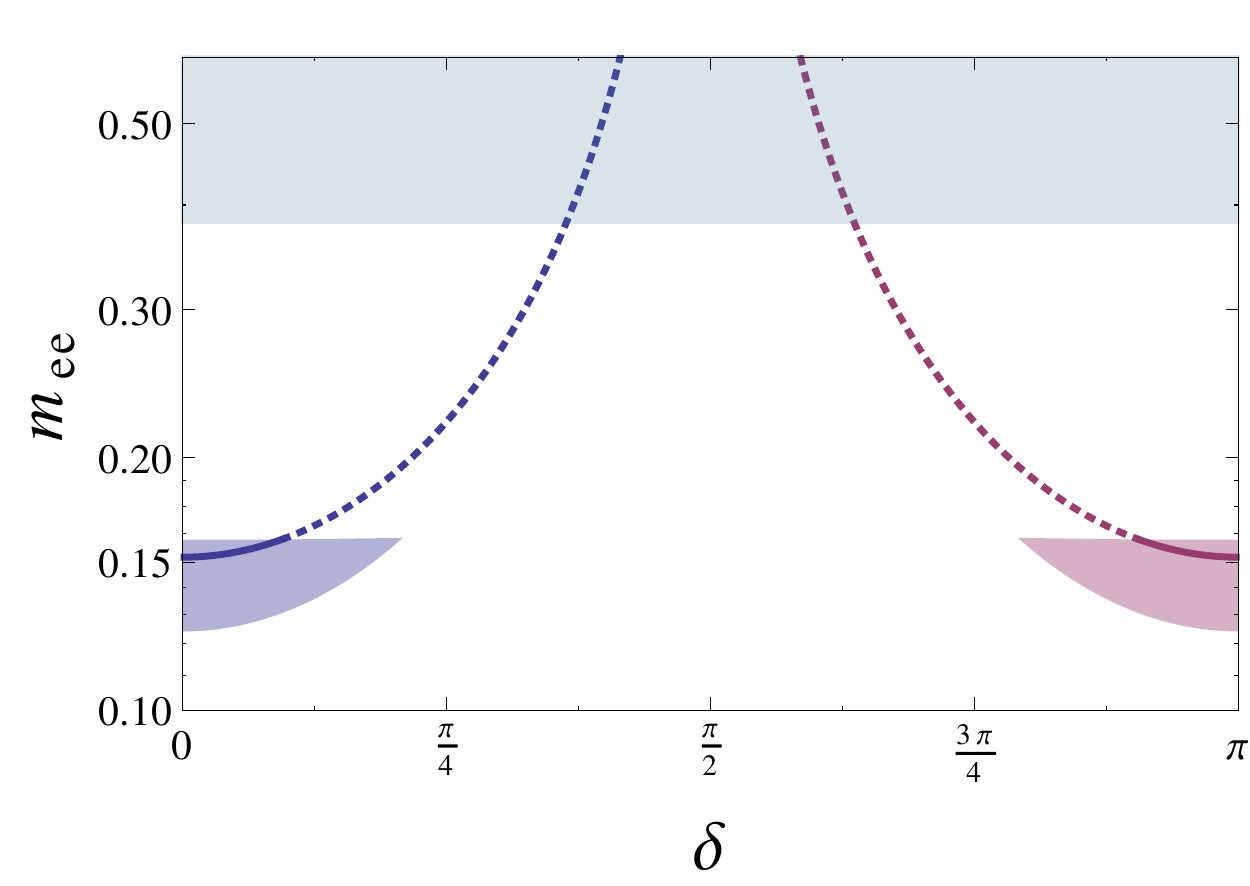}\\
\includegraphics[width=0.28\textheight]{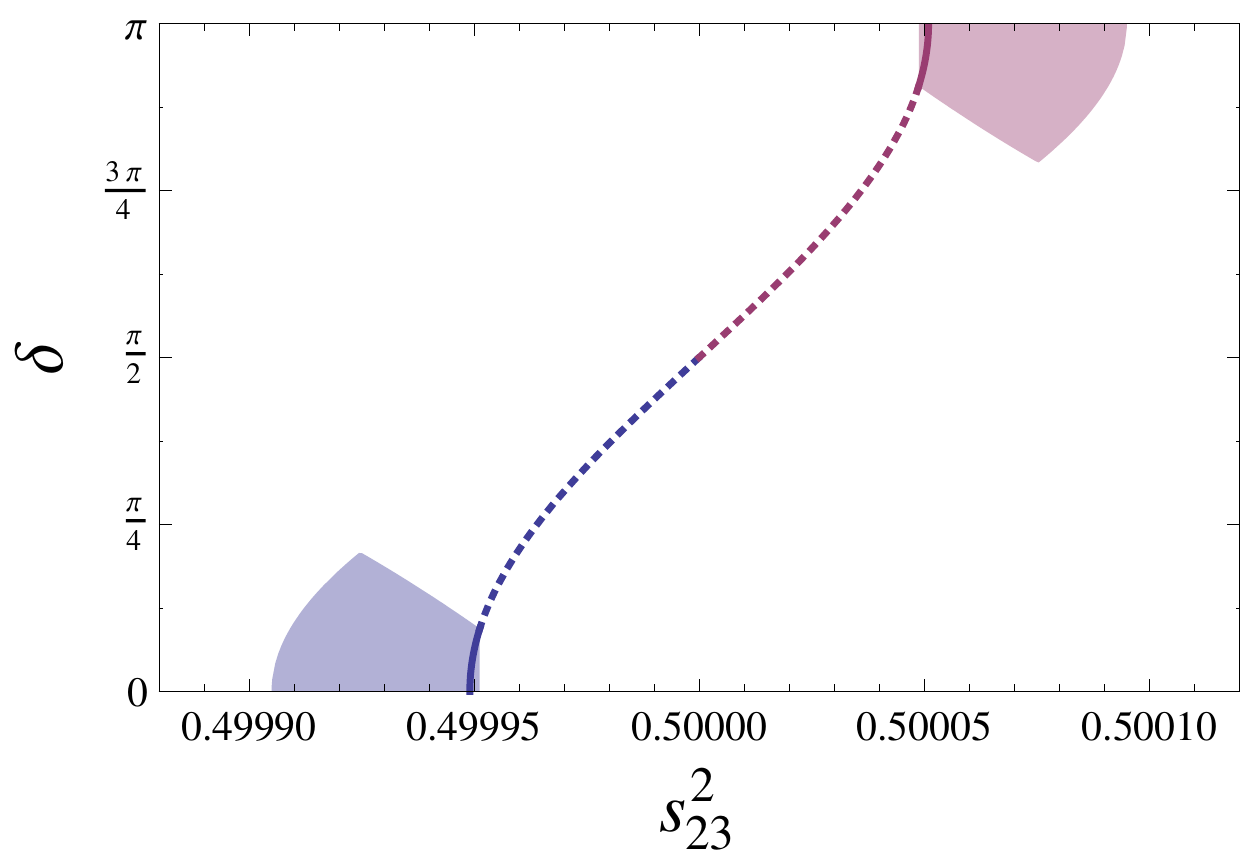} &
\includegraphics[width=0.28\textheight]{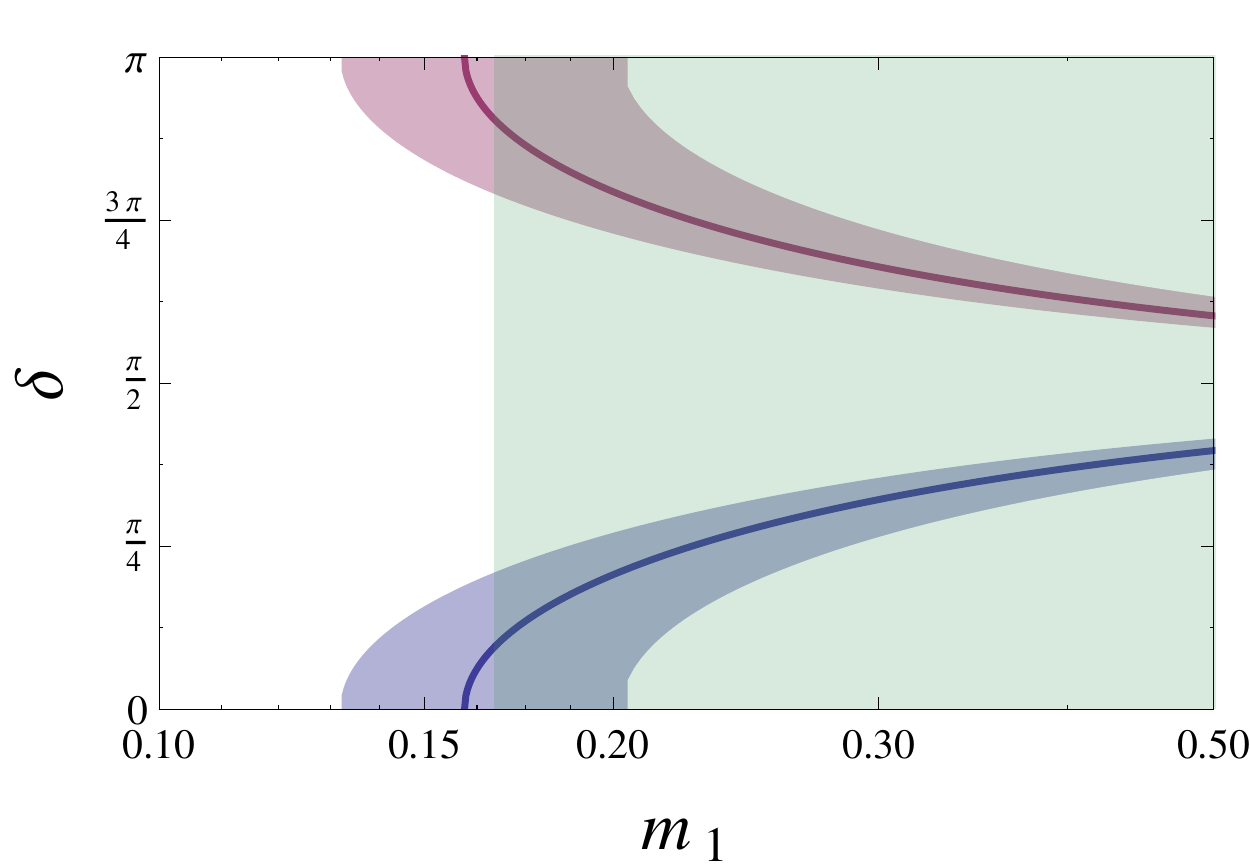}\\
\includegraphics[width=0.28\textheight]{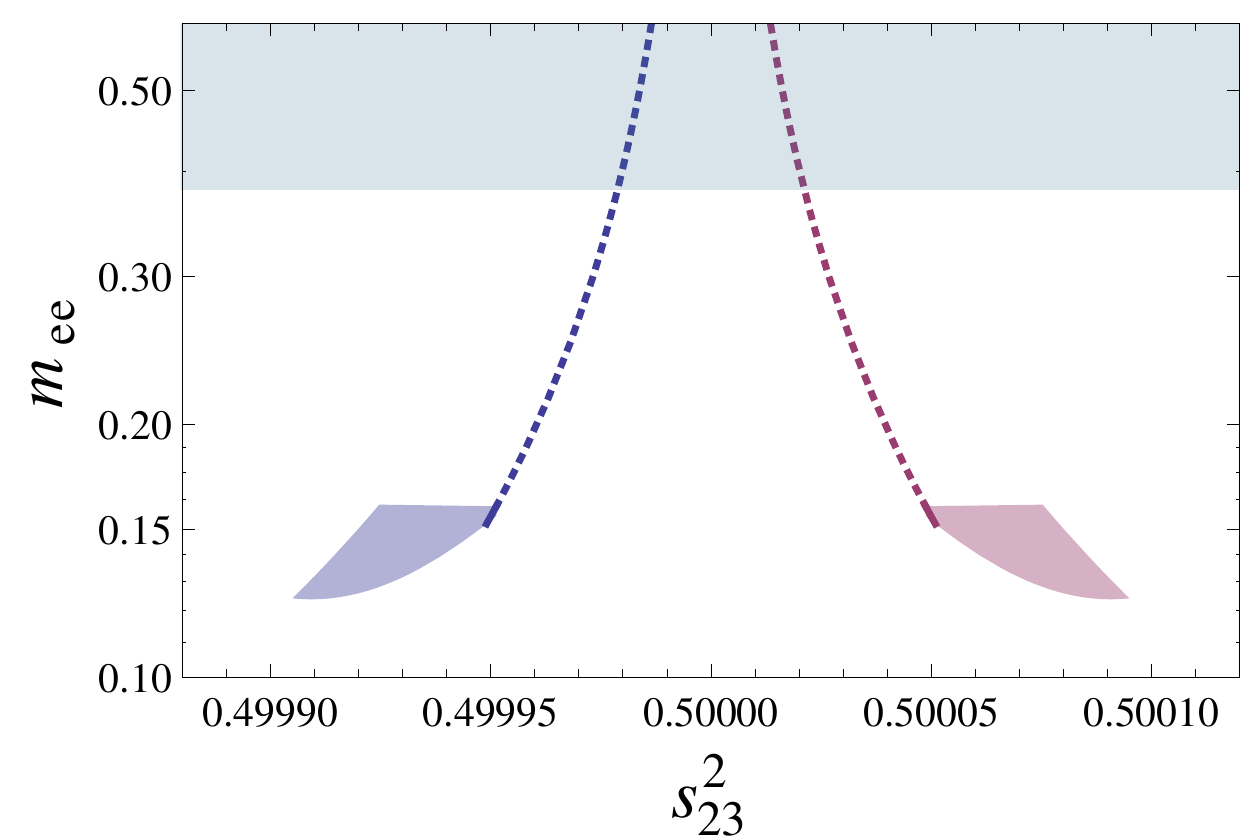}&
\includegraphics[width=0.28\textheight]{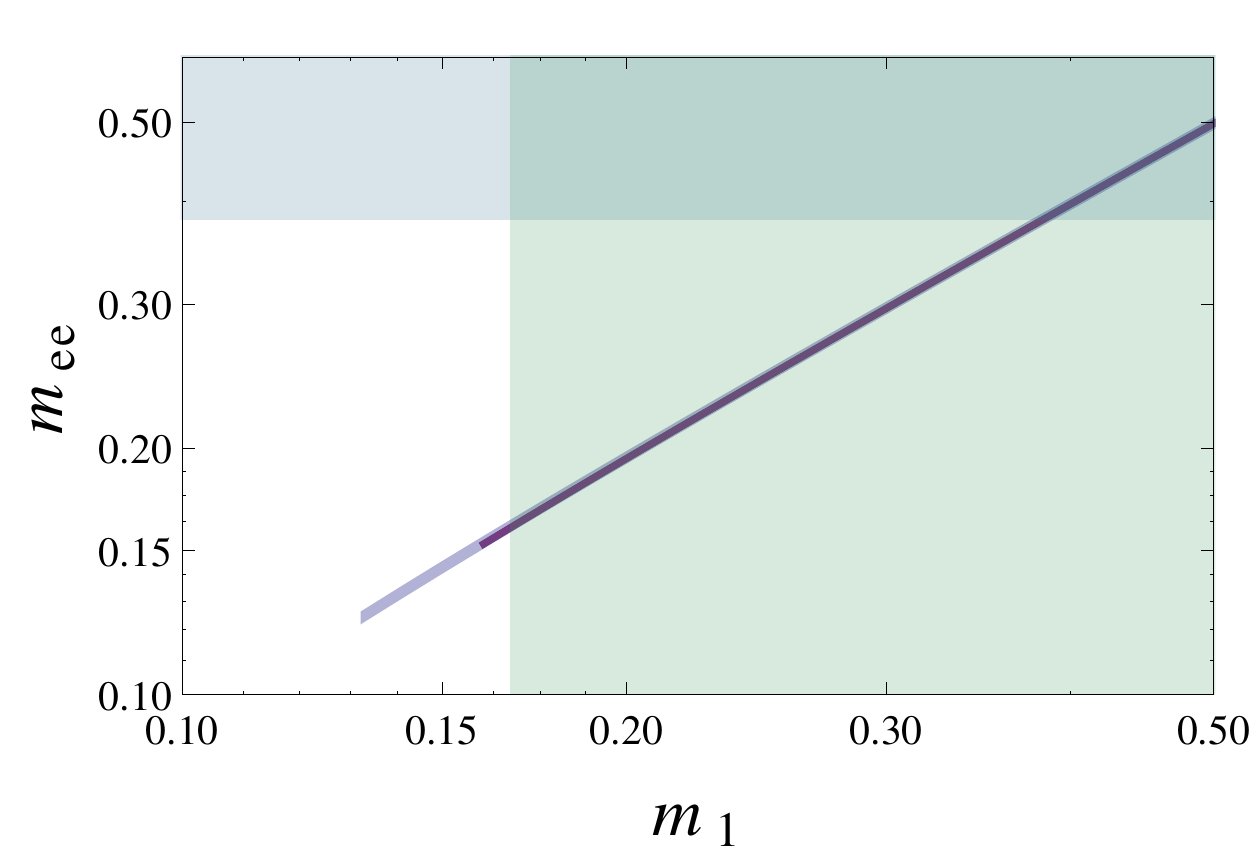}
\end{tabular}
\caption{\it The same as in Fig.~\ref{Cinverse}, but in the case of normal ordering.
}\label{Cnormal}\end{figure}

As the matrix  in \eq{C} maintains the same structure under conjugation, for $\delta \leftrightarrow -\delta$
the other observables are unchanged,
therefore in Figures 1 and 2 we need to display only the interval $[0,\pi]$ for $\delta$.
This matrix structure is also invariant under the exchange of the 2nd and 3rd rows and columns. 
This corresponds to the following exchange in the values of the observables:
\beq
\theta_{23} \leftrightarrow \frac\pi 2 -\theta_{23}~~~~{\rm and}~~~~ \delta\leftrightarrow \pi - \delta~.
\label{mutau}\eeq
Such relation can be clearly observed in the plots.
In the case of inverse ordering, illustrated in Fig.~\ref{Cinverse}, the $2-3$ mixing lies in the first (second) octant for $\cos\delta > 0$ ($<0$).
The curve $\delta(\theta_{23},p_a)$ is well described, up to small corrections of order $\Delta m^2_{21}/\Delta m^2_{31}$,
by the relation
\beq
\cos\delta \tan 2\theta_{23} \tan2\theta_{12} \sin\theta_{13} \simeq 1~.
\eeq
The mass spectrum can be  quasi-degenerate or hierarchical, with a lower bound $m_3\gtrsim 0.02$ eV, and $m_{ee}\simeq m_3$.
The upper bound on $\sum_i m_i$ implies $\sin^2\theta_{23} \lesssim 0.48$ with $\cos\delta\gtrsim 0.13$, or 
$\sin^2\theta_{23} \gtrsim 0.52$ with $\cos\delta\lesssim -0.13$.
In the case of normal ordering, illustrated in Fig.~\ref{Cnormal}, the $2-3$ mixing is extremely close to maximal, 
the mass spectrum is quasi-degenerate, with $m_i \gtrsim 0.15$ eV. 
As a consequence, values of $\delta$ close to $0$ or $\pi$ are predicted.
The equality $m_{ee}\simeq m_i$ holds in very good approximation.
The present upper bound on $\sum_im_i$ almost rules out the normal ordering case.
Our exact results are consistent with the analysis of the matrix in  \eq{C} performed in \Ref{Grimus:2004az}, where  the correlations between 
the $x_i$'s 
were understood by a sequence of careful approximations.

In order to identify a possible flavour symmetry responsible for the mass matrix structure in \eq{C},
it is useful to inspect the size of the parameters $a,b,c,d,$ that are needed to reproduce the experimental data. Note that, 
once the four input values
for the $p_a$'s are fixed, the mass matrix elements are fully determined as a function of one observable $x_j$. In Fig.~3 we show 
$m_{\alpha\beta}\equiv|(M_\nu)_{\alpha\beta}|$ as a function of $\sin^2\theta_{23}$, for $\alpha\beta = ee,e\mu,e\tau,\mu\tau$.
One observes that the pair $m_{ee}$ and $m_{\mu\tau}$ is quasi-degenerate in the whole range of $\theta_{23}$, 
while the pair  $m_{e\mu}$ and $m_{e\tau}$ becomes more and more degenerate as $\theta_{23}$ approaches the maximal value.
Note also that, in the case of inverse ordering, the two pairs cross each other for some special values of $\theta_{23}$, 
that actually lie in the present $1\sigma$ preferred regions.

\begin{figure}[t]
\begin{tabular}{cc}
\includegraphics[width=0.45\linewidth]{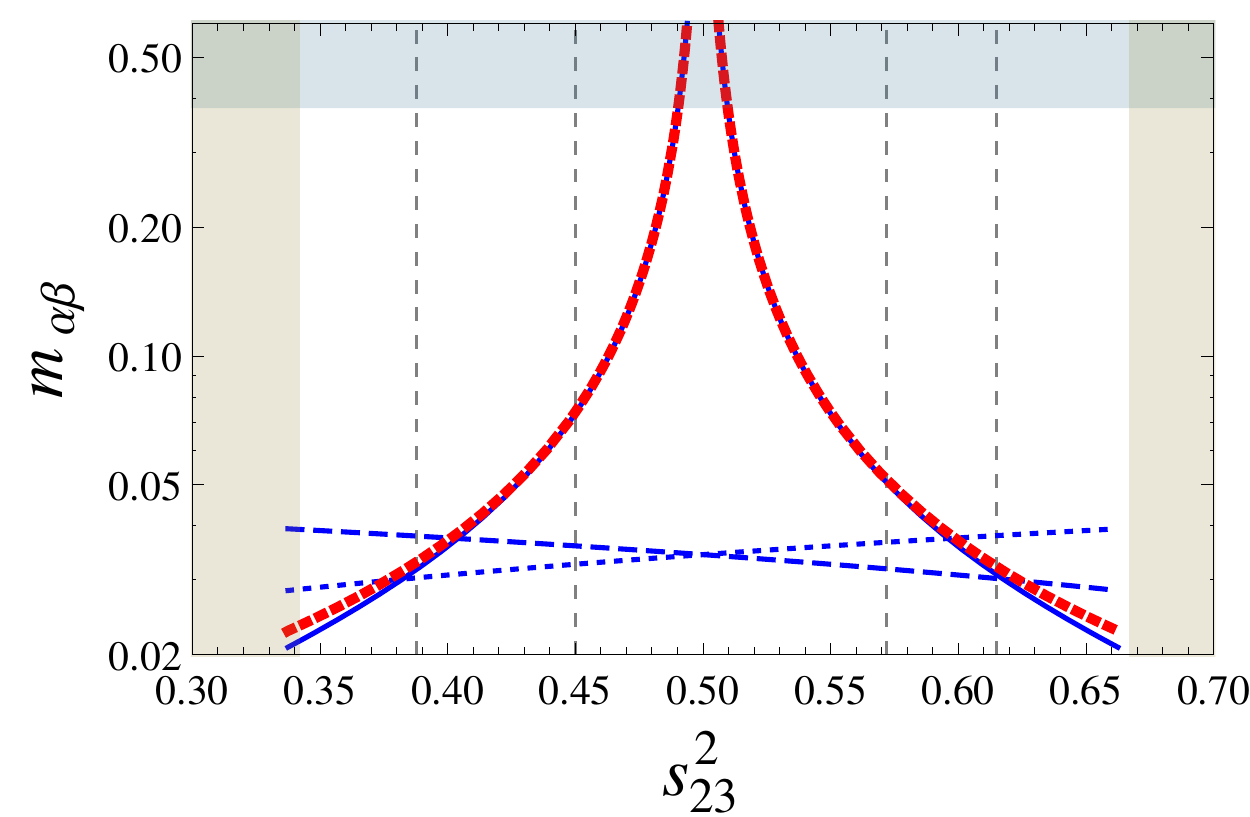}~~
\includegraphics[width=0.45\linewidth]{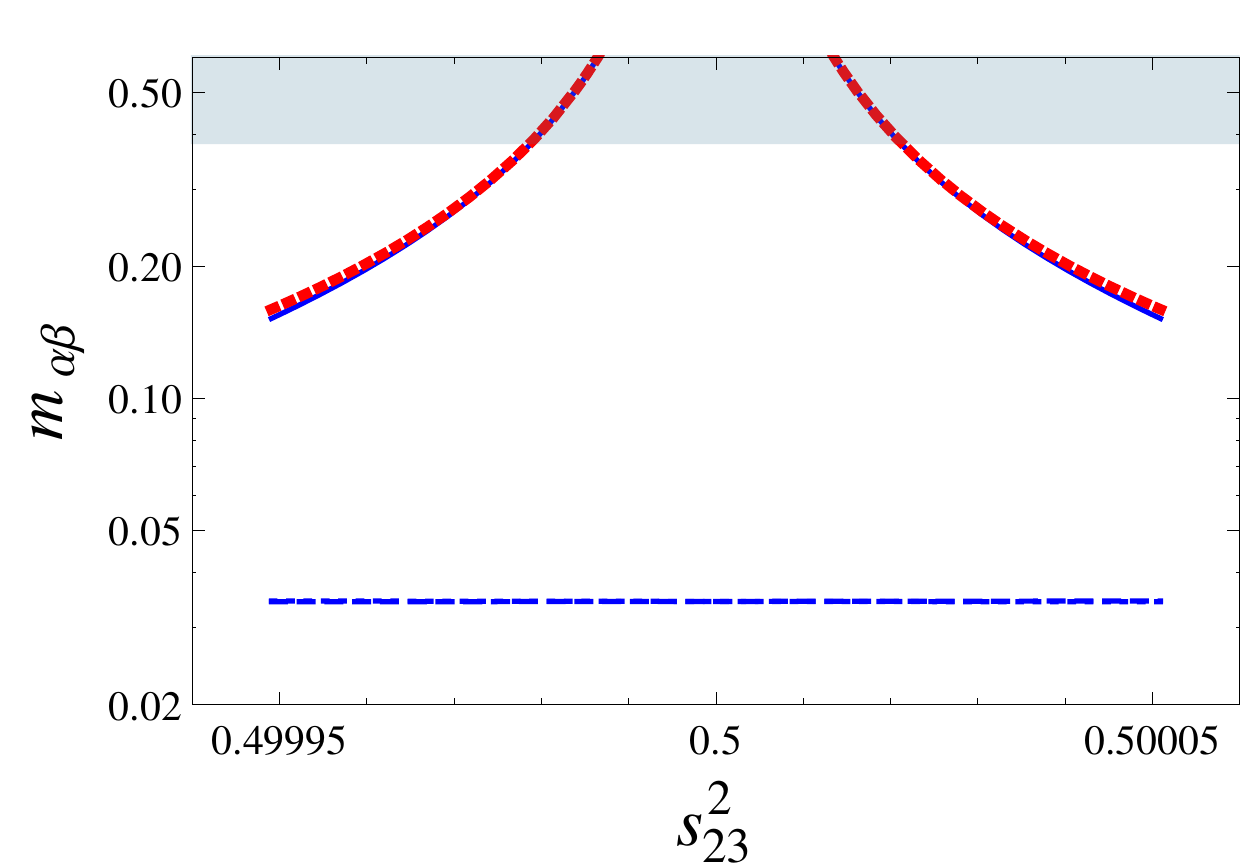}
\end{tabular}
\caption{\it
The absolute values $m_{\alpha\beta}$ (in eV) of the entries of the $M_\nu$ in \eq{C},  as a function of $\sin^2\theta_{23}$,
for the case of inverse (left-hand panel) and normal (right-hand panel) ordering. 
The thin solid blue line corresponds to $m_{ee}$, the thick dashed red line to $m_{\mu\tau}$, the thin dashed blue line to $m_{e\mu}$ and
the thin dotted blue line to $m_{e\tau}$. The shaded bands  are the same as in Fig.~\ref{Cinverse}.
The two pairs of vertical dashed lines limit the $1\sigma$ preferred region for $\theta_{23}$ in each octant.
}
\label{entriesC}\end{figure}

\section{A symmetry for zero and equal entries in the neutrino mass matrix}
\label{sec:dp3}

In the previous section we proposed a procedure to illustrate the phenomenological implications of a neutrino mass matrix depending on few parameters. 
From a theoretical point of view, one shall wonder whether the minimal viable matrix structures can be justified by a symmetry.

In this section we will focus on the matrix in \eq{C} in the limit $b=d$, 
since it corresponds to two values of $\sin^2\theta_{23}$ close to the presently preferred values in the first and in the second octant, as shown in Fig.~\ref{entriesC}. 
Taking
\beq
M_{\nu} = \begin{pmatrix}
a & b & c \\
b  & 0 & b \\
c  &b  & 0
\end{pmatrix}~,
\label{matrixq6}\eeq
and considering the $3\sigma$ allowed range for the parameters $p_a$'s, one finds two possible solutions:
\begin{align}
&\left\{\begin{aligned}
&\sin^2\theta_{23} = 0.40^{+0.02}_{-0.01}\\
&\cos\delta = 0.59^{+0.12}_{-0.14}\\
&m_\text{light} = m_3 = 0.037^{+0.001}_{-0.002}\,\text{eV}\\
&m_{ee} = 0.036^{+0.002}_{-0.001}\,\text{eV}
\end{aligned}\right. ~,
&
&\left\{\begin{aligned}
&\sin^2\theta_{23} = 0.62^{+0.03}_{-0.02}\\
&\cos\delta = -0.75^{+0.15}_{-0.12}\\
&m_\text{light} = m_3 = 0.0289^{+0.0002}_{-0.0001}\,\text{eV}\\
&m_{ee} = 0.0284^{+0.0000}_{-0.0001}\,\text{eV}
\end{aligned}\right. ~.
\label{matrixA}\end{align}
At first sight it may seem arduous to justify the vanishing of two entries of $M_\nu$ together with the equality of two other entries, at the same time maintaining 
the charged lepton mass matrix $M_e$ diagonal. 
Here we will illu\-strate the ingredients needed to achieve this result and build a representative flavour symmetry model.
We remark that, no matter how elaborate such model may be, it provides the sharp predictions in \eq{matrixA} for the observables.

The search for the appropriate flavour symmetry goes as follows:
\begin{itemize}

\item The equality of the matrix entries $M_{e\mu}$ and $M_{\tau\mu}$
requires that the two lepton doublets $l_e$ and $l_\tau$ transform together under the symmetry, that is, they belong to a dimension-two
irreducible representation (irrep) $2_l$, while $l_\mu$ is a flavour singlet $1_l$. In this way the two equal matrix entries can arise from a flavon doublet $\phi$, 
that is, a scalar field in a doublet irrep $2_\phi$, that acquires a vacuum expectation value (vev) in the direction $\langle\phi\rangle\propto (1,1)$.

\item The tensor product $2_l\times 2_l$ must lead to a non-zero diagonal entry, $M_{ee}=a$, and a vanishing one, $M_{\tau\tau}=0$.
Some simple group theory leads to the conclusion that this tensor product must contract with an additional flavon doublet $\phi'$, with 
$\langle\phi'\rangle\propto (1,0)$. Since $\phi$ and $\phi'$ must contribute separately in the sector $2_l\times 1_l$ and  $2_l\times 2_l$,
the flavour group must provide two different doublet irreps, $2_\phi\ne 2_{\phi'}$. 

\item Finally, since $M_{\mu\mu}$ vanishes, the tensor product $1_l\times 1_l$ should not be invariant under the flavour symmetry.

\end{itemize}
The simplest group that satisfies these properties is the order-twelve quaternion group $Q_6$, also known as the double-dihedral group $D'_3$,
that has two doublet irreps $2_1$ and $2_2$, plus four singlet irreps $1_i$, with in particular $1_3\times 1_3=1_4\times 1_4 = 1_2$, different from the
invariant singlet $1_1$. For the relevant group properties and conventions, we refer to \Ref{Blum:2007jz}, that provides useful tools for the analysis of 
(double-)dihedral symmetries in general.

Let us begin from the neutrino sector, writing the Majorana mass term as $y_{ij}^k l_i l_j \phi_k (h h/\Lambda)$, 
where $h$ is the Higgs doublet and $\phi_k$ are the  flavon fields that transform under $Q_6$
(we normalized $\phi_k$ to be dimensionless, that is, a scalar field over some cutoff scale). 
The lepton doublets and the flavons are assigned to the irreps of $Q_6$ in order to satisfy the above list of requirements,
\beq
\begin{pmatrix}l_e \\ l_\tau \end{pmatrix} \sim 2_2,~~
l_\mu\sim 1_3,~~
\begin{pmatrix}\phi_{1u}\\ \phi_{1d}\end{pmatrix}\sim 2_1,~~
\begin{pmatrix}\phi_{2u}\\ \phi_{2d}\end{pmatrix}\sim 2_2.
\eeq
Then, the $Q_6$-invariant Lagrangian has the form
\beq
-{\cal L}_\nu = \left[ 
y_{e\tau} l_e l_\tau + \frac 12 y_{ee} (l_e l_e \phi_{2u} + l_\tau l_\tau \phi_{2d}) + y_{e\mu}  (l_e\phi_{1u}+l_\tau\phi_{1d})l_\mu
\right]\frac{hh}{\Lambda} +h.c. ~.
\eeq
Taking $\langle\phi_1\rangle=c_1(1,1)$ and $\langle\phi_2\rangle=c_2(1,0)$, one obtains the neutrino mass matrix in \eq{matrixq6},
with $a = y_{ee}c_2v^2/\Lambda$, $b=y_{e\mu}c_1v^2/\Lambda$ and $c = y_{e\tau} v^2/\Lambda$.

The model works as long as $M_e$ is diagonal in the same basis. The charged lepton mass term can be written as
 $\lambda^k_{ij}l_ie^c_j\varphi_k h^*$, where $e_j^c$ are the charged lepton singlets and $\varphi_k$ the flavons,
 in general different from those in the neutrino sector.
With the assignment
\beq
\begin{pmatrix}\mu^c \\ e^c \end{pmatrix} \sim 2_1,~~
\tau^c \sim 1_3,~~
\begin{pmatrix}\varphi_{1u}\\ \varphi_{1d}\end{pmatrix}\sim 2_1,~~
\begin{pmatrix}\varphi_{2u}\\ \varphi_{2d}\end{pmatrix}\sim 2_2,
\eeq
the $Q_6$-invariant Lagrangian reads
\bea
-{\cal L}_e &=& \left[ 
\lambda_e (l_e e^c \varphi_{1d}- l_\tau \mu^c\varphi_{1u}) + 
\lambda_\mu l_\mu (\mu^c\varphi_{2u}+e^c\varphi_{2d}) \right. \nonumber\\
&+& \left.
\lambda_\tau (l_\tau \varphi_{1d} + l_e\varphi_{1u})\tau^c
\right]h^* +h.c. ~.
\eea
For $\langle\varphi_1\rangle=c'_1(0,1)$ and $\langle\varphi_2\rangle=c'_2(1,0)$, the charged lepton mass matrix is diagonal,
with $m_e = \lambda_e c'_1 v$, $m_\mu = \lambda_\mu c'_2 v$ and $m_\tau=\lambda_\tau c'_1 v$.

This completes the prove of existence of a minimal model leading to \eq{matrixq6}. 
Note that the flavour group $Q_6$ is entirely broken by the flavon vevs, both in the neutrino and in the charged lepton sector.

\section{Off-diagonal plus universal neutrino masses}
\label{sec:zeeplusd}

Here we propose another, remarkably simple structure of the neutrino mass matrix that depends on five physical parameters and predicts the preferred, non-maximal $\theta_{23}$. We will justify this structure with a minimal flavour symmetry at the end of this section.

\begin{figure}[t]
\begin{tabular}{cc}
\includegraphics[width=0.45\textwidth]{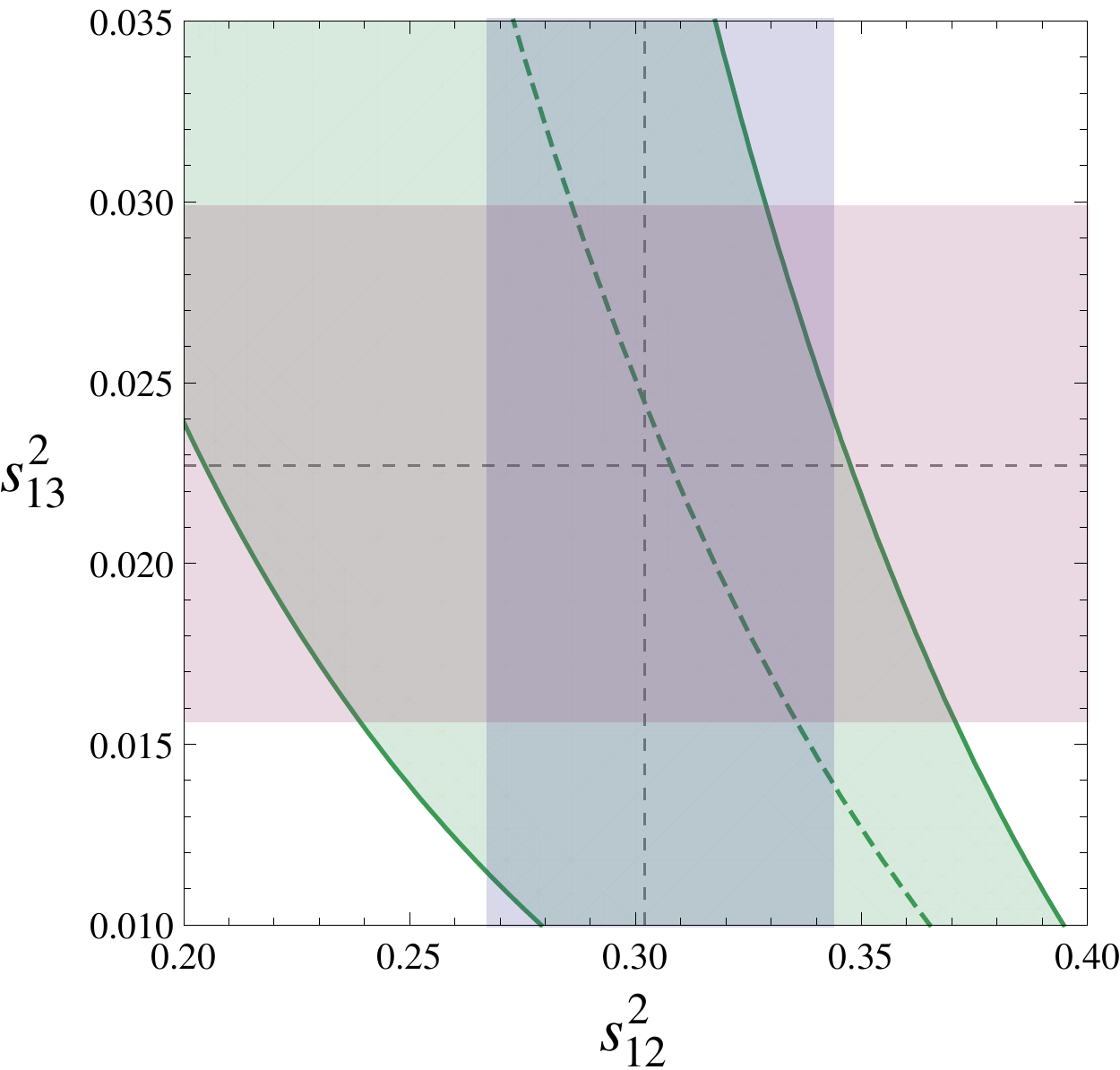}&
\includegraphics[width=0.45\textwidth]{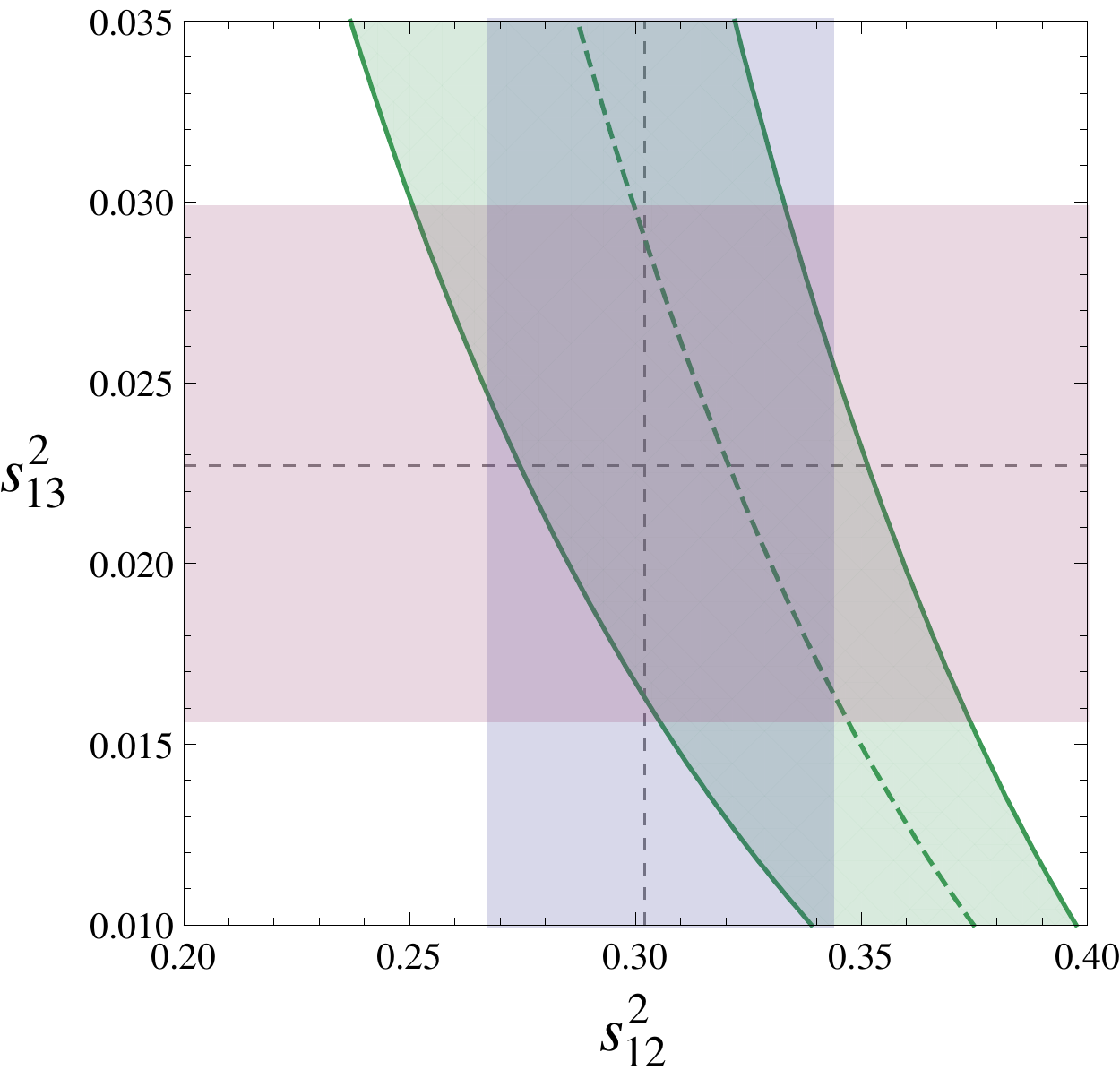}
\end{tabular}
\caption{\it 
The predicted value of $\sin^2\theta_{23}$ as a function of $\sin^2\theta_{12}$ and $\sin^2\theta_{13}$, for the off-diagonal neutrino mass matrix of  
\eq{mzee}. The green shaded band corresponds to the $1\sigma$ range for $\sin^2\theta_{23}$ in the first octant (left-hand panel) and in the second
octant (right-hand panel).
The vertical (horizontal) band corresponds to the $3\sigma$ range for $\sin^2\theta_{12}$ ($\sin^2\theta_{13}$).
Dashed lines are the best fit values of the mixing angles.}
\label{fig:zee}
\end{figure}

Our starting point is the so-called Zee mass matrix  \cite{Zee:1980ai,Wolfenstein:1980sy},
with only off-diagonal entries, 
\begin{equation}
M^\text{off}_{\nu} = 
\begin{pmatrix}
0 & a & b \\
a & 0 & c \\
b & c & 0
\end{pmatrix},
\label{mzee}\end{equation}
in the basis where $M_e$ is diagonal.
The phases of $a$, $b$ and $c$ can be rotated away, thus there is no CP-violation 
and the leptonic mixing is described by a real orthogonal matrix $R$, with $M_\nu^\text{off} = R ~\text{diag}(\lambda_1,\lambda_2,\lambda_3)\, R^T$. 
A simple way to extract a constraint on the mixing angles is to regard 
the three zero diagonal entries as a homogeneous system of three linear equations, 
\begin{equation}
\sum_j Z_{ij} \lambda_j = 0,~~i=1,2,3,
\end{equation}
where $Z_{ij}=R_{ij}^2$. 
In order to have a non-trivial solution, the matrix $Z$ must be singular. The equation $\det Z =0$ 
establishes a relation between the mixing angles, 
\begin{equation}
\frac{1}{\tan(2\theta_{12})}\frac{1}{\tan(2\theta_{23})} \frac{2(1-2\sin^2\theta_{13})}{\sin\theta_{13}(1-3\sin^2\theta_{13})} = \pm 1~,
\label{eq:zee}\end{equation}
where the positive (negative) sign corresponds to $\theta_{23}<\pi/4$ ($\theta_{23}>\pi/4$) and $\delta=0$ ($\delta=\pi$).
Not surprisingly, this equation has been already derived in various ways in the literature (see e.g. Refs.~\cite{Jarlskog:1998uf,Chen:2005jm}).
Indeed, the corresponding prediction for $\theta_{23}$  fulfills with great accuracy the most recent oscillation data,
as shown in Fig.~\ref{fig:zee}.
Taking the best fit values and the $3\sigma$ ranges for $\theta_{12}$ and $\theta_{13}$ from \Ref{GonzalezGarcia:2012sz}, \eq{eq:zee} implies 
\begin{equation}
\sin^2\theta_{23} = 
0.42^{+0.02}_{-0.04} {\rm~~or~~} 0.58^{+0.04}_{-0.02}~.
\end{equation}
It is remarkable that 
these intervals roughly coincide with 
the $1\sigma$ preferred regions for $\theta_{23}$.

The Zee mass matrix is known to be unable to reproduce all current data \cite{Jarlskog:1998uf,Frampton:1999yn,He:2003ih}: 
it was shown that, if one fits the two neutrino mass squared differences, then
the resulting mixing pattern is close to bi-maximal, which rules out the model. 
This is simply because the three parameters  $a$, $b$ and $c$ are insufficient to accommodate the values of the four precisely measured observables $p_a$'s.
In fact, the three zeroes on the diagonal imply \eq{eq:zee} as well as two relations between the eigenvalues,
\beq
\lambda_1+\lambda_2+\lambda_3=0 ~,~~~~~\lambda_1(\cos^2\theta_{12}-\tan^2\theta_{13})=\lambda_2(\tan^2\theta_{13}-\sin^2\theta_{12})~.
\eeq
Given the measured values of the mixing angles, the mass squared differences $\Delta m^2_{ij}=\lambda_i^2-\lambda_j^2$
are incompatible with data.

Now, the perfect prediction for $\theta_{23}$ makes us wonder whether the lepton mixing pattern is indeed determined by an 
off-diagonal mass term, while the neutrino mass eigenvalues are not. Then, the most natural extension of \eq{mzee} is
the addition of a universal mass term, 
\begin{equation}
M_{\nu} = 
\begin{pmatrix}
0 & a & b \\
a & 0 & c \\
b & c & 0
\end{pmatrix}
+
d \begin{pmatrix}
1 & 0 & 0 \\
0 & 1 & 0 \\
0 & 0 & 1
\end{pmatrix}~,
\label{offu}\end{equation}
with $d=|d|e^{i \chi}$ complex in general. This modification does not affect the mixing matrix, as long as $a$, $b$ and $c$ are real (or have the same phase). 
The neutrino mass eigenvalues are given by
\begin{equation}
m_i = |\lambda_i + d|~,
\end{equation}
and one is able to accommodate the two neutrino mass squared differences. 
In particular, we obtain a condition on the mixing angles that determines the ordering of the neutrino mass spectrum: 
\begin{equation}
{\rm normal~for~~}s_{12}^2 < \frac{t_{13}^2+1}{3}~,~~~~~{\rm inverse~for~~}s_{12}^2 > \frac{t_{13}^2+1}{3} ~.
\end{equation}
The experimental values of $\theta_{12}$ and $\theta_{13}$ prefer normal ordering, with the inverse one only marginally allowed at $3\sigma$.

\begin{figure}[t]
\begin{tabular}{cc}
\includegraphics[width=0.45\textwidth]{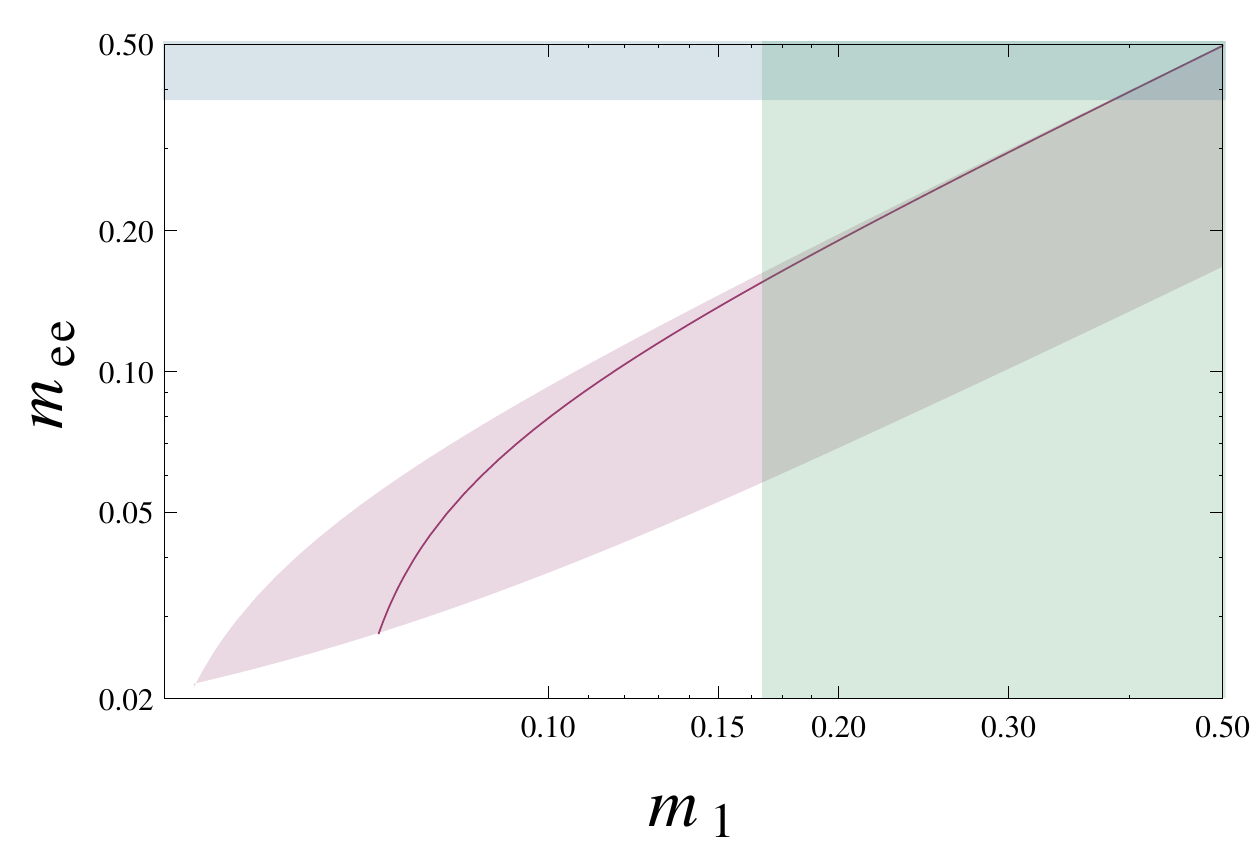}&
\includegraphics[width=0.45\textwidth]{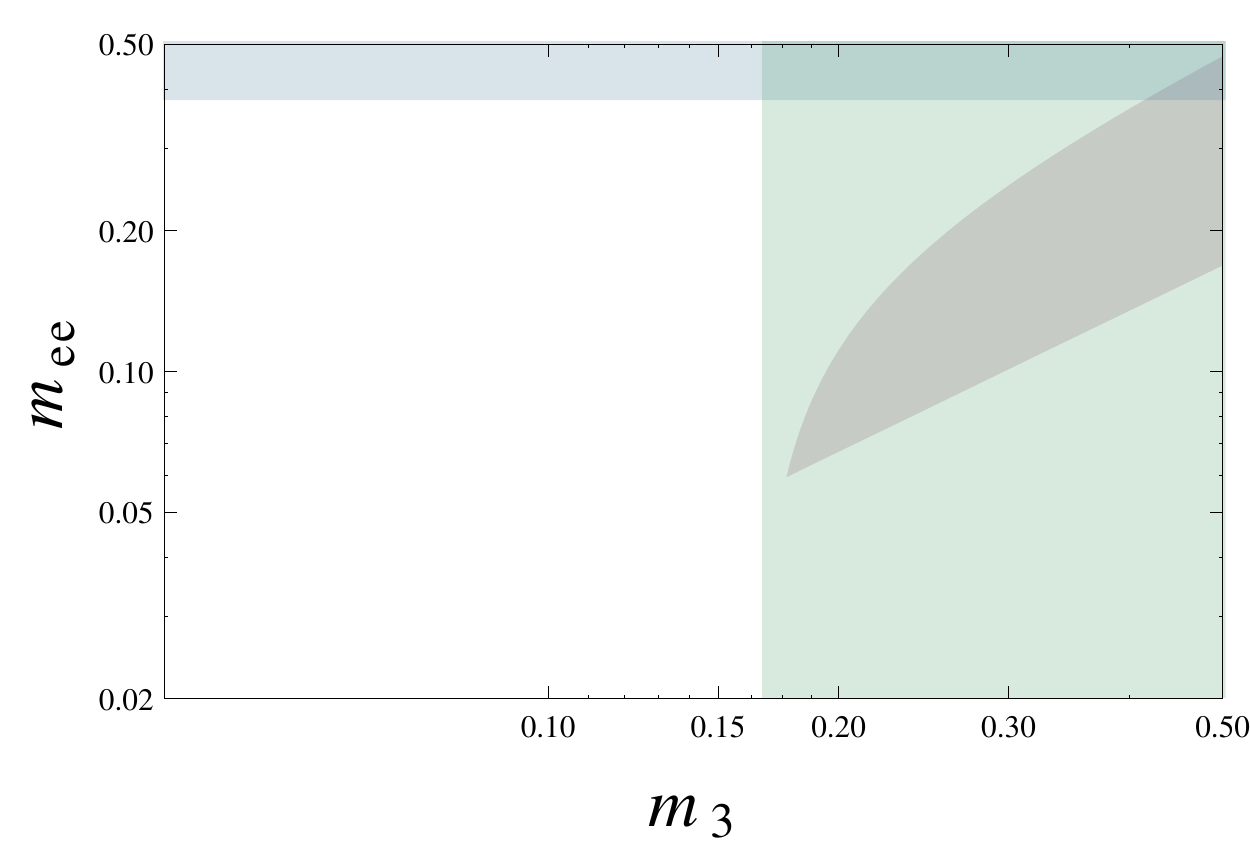}
\end{tabular}
\caption{\it 
The effective mass $m_{ee}$ as a function of $m_{\text{light}}$ for the neutrino mass matrix of \eq{offu}, in the case of normal (left-hand panel) 
and inverse (right-hand panel) ordering. The solid line corresponds to the best fit input parameters $p_a$'s, while the purple shaded regions 
correspond to their $3\sigma$ allowed  range. 
The green (blue) band is disfavoured by the cosmological upper bound on $\sum m_i$ (by  $0\nu2\beta$-decay searches).}
\label{fig:zee-mee-mlight}
\end{figure}

Note that the $M_\nu$ in \eq{offu} depends on five physical parameters, but it behaves in a singular manner with respect to our general analysis 
of section \ref{preds}:
the four input parameters $p_a$'s fix $a$, $b$, $c$ (or equivalently the three $\lambda_i$'s) and 
the product $|d|\cos\chi$. Two output parameters are 
automatically determined by \eq{eq:zee}:
$\theta_{23}$ and $\delta$ (equal to $0$ or $\pi$).  
As $d$ is in general complex, there is CP-violation, but only in the form of Majorana phases. 
The other two output parameters, $m_\text{light}$ and $m_{ee}$, depend on $|d|$ and $\cos\chi$ separately,
and thus there is a non-trivial correlation between them, $m_{ee}=m_{ee}(m_\text{light},p_a)$, that is illustrated in Fig.~\ref{fig:zee-mee-mlight}.
Since the product $|d| \cos\chi$ is fixed by the $p_a$'s, 
there is a lower bound on $m_{ee}\equiv |d|$, which translates into a lower bound  on $m_\text{light}$ too. 
In the case of normal ordering, taking into account the $3\sigma$ uncertainties on the $p_a$'s, we obtain
\begin{equation}
m_1 \gtrsim 4.3\times 10^{-2}\,\text{eV},\qquad m_{ee} \gtrsim 2.1\times 10^{-2}\,\text{eV}.
\end{equation}
The inverse ordering is disfavoured by the cosmological bound on $\sum m_i$.

Let us show that the matrix in \eq{offu} can be motivated by a minimal flavour symmetry. Clearly, the three lepton doublets
shall transform in a triplet irrep of the flavour group. The smallest group with such an irrep is $A_4$, the alternating group of 4 objects
(we adopt the conventions in \Ref{Chen:2005jm}).
The universal mass term is precisely the unique $A_4$-invariant contribution to $M_\nu$, while
the off-diagonal mass term is generated by an $A_4$-triplet flavon field $(\phi_1,\phi_2,\phi_3)$:
\beq
-{\cal L}_\nu = \left[\frac 12 y_\text{uni} \sum_\alpha l_\alpha l_\alpha + y_\text{off} (\phi_1 l_\mu l_\tau + l_e \phi_2 l_\tau + l_e l_\mu \phi_3)\right] 
\frac{hh}{\Lambda} + h.c. ~.
\eeq
Note that, in order for $a$, $b$ and $c$ to have the same phase, the three vevs $\langle\phi_i\rangle$ should be real, that is the case
as long as the flavon potential does not break CP. 
The other requirement is to keep  $M_e$ diagonal, with three different eigenvalues. This is promptly obtained
as soon as the charged lepton isosinglets $e^c_i$ also transform as an $A_4$-triplet. Introducing flavons $\varphi'$ and $\varphi''$ 
in the irreps $1'$ and $1''$ of $A_4$, respectively, one obtains
\bea
-{\cal L}_e &=& \Big[y_1 \sum_\alpha l_\alpha e^c_\alpha 
+ y_{1'}\varphi'(l_e e^c + \omega l_\mu \mu^c + \omega^2 l_\tau \tau^c)  \nonumber\\
&+&  y_{1''}\varphi''(l_e e^c + \omega^2 l_\mu \mu^c + \omega l_\tau \tau^c)
{\Big{]}} h^* + h.c. ~,
\eea
where $\omega=\exp(2\pi i/3)$. This leads to $M_e=\text{diag}(m_e,m_\mu,m_\tau)$ as desired.
Note that $A_4$ is broken to nothing in the neutrino sector, because one needs $\langle\phi_1\rangle\ne \langle\phi_2\rangle \ne \langle\phi_3\rangle\ne 0$,
while in the charged lepton sector a $Z_2\times Z_2$ subgroup is preserved.

\section{Minimal 1+2 lepton flavour symmetry \label{1+2}}

In the sections \ref{sec:dp3} and \ref{sec:zeeplusd} we started from predictive mass matrix structures and built a flavour model to justify them.
Here we follow the opposite path: we will construct a flavour model, with the minimal possible number of technical ingredients, and study the corresponding correlations between the observables.

Let us hypothesize that the three families of lepton doublets $l_i$ as well as the three charged lepton singlets $e^c_i$ transform in a dim-one and a dim-two irreps of
a flavour symmetry group:
\beq
l_i \sim (1_l + 2_l)_i~,~~~~~ e^c_i \sim (1_{e^c} + 2_{e^c})_i~.
\eeq
In order to generate a non-zero mixing between the flavour singlet family and the doublet ones, it is mandatory to introduce a flavon in a dim-two irrep, $\phi\sim 2_\phi$,
that acquires a non-zero vev. 
In the search for a minimal theory of lepton flavour, it is meaningful to ask whether a viable model exists, 
that contains only flavons of this type, 
and what are the associated predictions.

The possible contributions to the lepton mass matrices are given by the irrep tensor products that
are invariant under the flavour group:
\bea
M_\nu &\sim& \left(\ba{c|cc}
(1_l \times 1_l)_1 & \multicolumn{2}{c}{(1_l\times 2_l\times 2_{\phi^\nu})_1}\\ \hline
\dots & \multicolumn{2}{c} {(2_l\times 2_l)_1+(2_l\times 2_l\times 2_{\phi^\nu})_1}\\
\ea\right)~,\label{mnu1+2}\\
M_e &\sim& \left(\ba{c|cc}
(1_l \times 1_{e^c})_1 & \multicolumn{2}{c}{(1_l\times 2_{e^c}\times 2_{\phi^e})_1}\\ \hline
(2_l\times1_{e^c}\times2_{\phi^e})_1 & \multicolumn{2}{c} {(2_l\times 2_{e^c})_1+(2_l\times 2_{e^c}\times 2_{\phi^e})_1}\\
\ea\right)~,\label{me1+2}
\eea
where the subscript $1$ denotes the invariant component in the tensor products, if any.
Note that we included at most one power of the flavon irrep; this is the whole story if the theory is renormalizable and the flavon fields carry also gauge indexes, that is
to say, $\phi^\nu$ is a Higgs triplet, and $\phi^e$ a Higgs doublet. Alternatively, one can treat the flavons as gauge singlets; in this case
higher dimensional operators, suppressed by a cutoff scale $\Lambda$,
may contain several powers of the flavons, leading to more complicated flavour structures; here we will neglect these corrections of order $(\langle \phi \rangle/\Lambda)^n$.

Let us identify the simplest viable mass matrices with the structure given in \eq{mnu1+2} and \eq{me1+2}. The minimal possibility is 
to employ a unique dim-two irrep
($2_l\equiv 2_{e^c} \equiv 2_\phi \equiv 2$), and to assume that the tensor product $2\times 2$ contains a singlet, while $2\times 2\times 2$ does not (to avoid an additional, unnecessary 
invariant). This is the case for the two non-abelian order-eight groups $D_4$ and $Q\equiv D'_2$, that satisfy $2\times 2 = \sum_{i=1}^4 1_i$, where $1_1$ is the invariant singlet
(we use conventions as in Ref.~\cite{Frigerio:2004jg}).
In the case of the quaternion group $Q$, the invariant component in $2_l\times 2_l$ is antisymmetric, therefore the corresponding contribution to $M_\nu$ vanishes; we will see 
that such contribution is needed to accommodate the data, thus we will rather focus on the dihedral symmetry $D_4$.
In this case $(2_l\times 2_l)_1 = l_1 l_1 + l_2 l_2$ and  the lepton mass matrices take the form
\beq
M_\nu = \left(\ba{ccc}
a & b & b' \\ 
\dots & c & 0 \\
\dots & \dots & c
\ea\right)~,~~~~~
M_e = \left(\ba{ccc}
0 & B & B' \\
D & C & 0 \\
D' & 0 & C 
\ea\right)~,\label{memnu}
\eeq
where the diagonal entries are independent from the flavons, while the off-diagonal ones are linear in the flavons.
According to our minimality tenet, we put to zero the $11$-entry of $M_e$ by choosing $1_l\times 1_{e^c} \ne 1_1$. 
Denoting the flavon vevs by $\langle \phi^\nu \rangle = (\phi^\nu_u~\phi^\nu_d)^T$ and 
$\langle \phi^e \rangle = (\phi^e_u~\phi^e_d)^T$, one has $b/b' = \pm (\phi^\nu_u/\phi^\nu_d)^{\pm1}$, $ B/B' = \pm(\phi^e_u/\phi^e_d)^{\pm1}$ and 
$D/D'=\pm(\phi^e_u/\phi^e_d)^{\pm1}$, where the signs are not correlated and depend on the choice of the irreps $1_l$ and $1_{e^c}$.

The physical mixing matrix $U_{PMNS}$ is determined as usual,
\beq
M_\nu = U^*_\nu d_\nu U_\nu^\dag ~,~~~~~ M_e M_e^\dag = U_e^* d_e^2 U_e^T~,~~~~~U_{PMNS} = U_e^\dag U_\nu~,
\label{pmns}\eeq
where $d_\nu$ and $d_e$ are diagonal matrices containing the neutrino and charged lepton mass eigenvalues.
Let us remark that the vev of a doublet $\phi$ breaks $D_4$ completely, unless it is aligned in the directions $(1,\pm1)$, $(1,0)$ or $(0,1)$: in these cases
a $Z_2$ subgroup is preserved.
We will consider only  these vev alignments, that are protected by the residual symmetry and correspond to a minimal number of parameters in the mass matrices.
In the case $b=0$ (or $b'=0$), $M_\nu$ is diagonalized by a single rotation of angle $\theta_\nu$; the case $b=\pm b'$ can be reduced to the previous one by a
$\pi/4$ rotation in the flavour doublet sector. Similarly, for $B=D=0$ (or $B=D'=0$, etc.), $M_eM_e^\dag$ is diagonalized by a unique rotation $\theta_e$, while
for $B=\pm B'$ and $D=\pm D'$ an additional maximal rotation is needed. 
If this maximal mixing is present (or absent) in both $M_\nu$ and $M_eM_e^\dag$, it cancels out in $U_{PMNS}$ and
it is straightforward to check that one entry of $U_{PMNS}$ vanishes, at odds with experimental data \cite{GonzalezGarcia:2012sz}. 
Thus, data indicate that $\phi^\nu$ and $\phi^e$ should be two different flavons, one acquiring a vev in the direction $(1,0)$ or $(0,1)$, and the other
in the direction $(1,\pm 1)$.

The general idea to break a flavour symmetry group $G$ to two different subgroups $G_\nu$ and $G_e$ in the neutrino and in the charged lepton sector, respectively, is not new:
it was mostly employed to explain the tri-bi-maximal mixing pattern, in models where the three lepton doublets transform in a triplet representation of $G$
\cite{Altarelli:2005yx}.
Here we have shown that, in the context of $1+2$ flavour models, the minimal viable possibility is to take $G_\nu=Z_2$ and $G_e=Z'_2$.

Let us derive the predictions of this scenario.
Following the above considerations, in \eq{memnu} we take $b'=0$, $B'=B$ and $D'=-D$ (there are a few other equivalent assignments). 
Then, one can check that $M_\nu$ and $M_eM_e^\dag$ depend on 6 independent absolute values of the matrix entries, plus 2 physical phases,
therefore the 6 lepton masses, the 3 mixing angles and the 3 CP-violating phases are correlated, as we now describe.
One finds
\beq
U_\nu = \left(\ba{ccc}
\cos\theta_\nu e^{-i\alpha} & 0 & \sin\theta_\nu e^{-i\alpha} \\
-\sin\theta_\nu e^{i\alpha} & 0 & \cos\theta_\nu e^{i\alpha} \\
0 & 1 & 0
\ea\right)~,~~~~~ d_\nu = \text{diag}(m_-,c,m_+) ~,
\eeq
where $\alpha= \arg(ab^*+bc^*)/2$, $\tan 2\theta_\nu = 2|ab^*+bc^*|/(|c|^2-|a|^2)$,
$|m_\pm|^2 = |b|^2 + |a|^2/2+|c|^2/2 \pm\sqrt{(|c|^2-|a|^2)^2/4+|ab^*+bc^*|^2}$.
Note that the two independent mass squared differences $\Delta m^2_{ij}$ can be accommodated independently from the value of 
two parameters $\alpha$ and $\theta_\nu$, that enter in $U_{PMNS}$.
In the charged lepton sector one has
\beq
U_e = \left(\ba{ccc}
\cos\theta_e  & \sin\theta_e e^{-i\beta} & 0\\
-\dfrac{\sin\theta_e e^{i\beta}}{\sqrt 2} &  \dfrac{\cos\theta_e}{\sqrt 2} & -\dfrac{1}{\sqrt2} \\
-\dfrac{\sin\theta_e e^{i\beta}}{\sqrt 2} &  \dfrac{\cos\theta_e}{\sqrt 2} & \dfrac{1}{\sqrt 2} 
\ea\right)~,~~~ d_e = \text{diag}(0,m_\mu,m_\tau) ~,
\eeq
with $\beta=\arg(BC^*)$, $m_\mu^2=2|B|^2+|C|^2$, $m_\tau^2=2|D|^2+|C|^2$ and $\tan\theta_{e}=\sqrt{2}|B/C|$. Note that $m_e$ vanishes,
which is a very good approximation, and the two masses $m_\mu$ and $m_\tau$ are independent from the two parameters $\theta_e$ and $\beta$,
that enter in $U_{PMNS}$ 
(if the assignment of $m_\mu$ and $m_\tau$ were interchanged,
$\theta_e$ would be forced to lie close to $\pi/2$ and the observed mixing angles could not be reproduced).

\begin{figure}[t]
\bec
\includegraphics[width=0.60\textwidth]{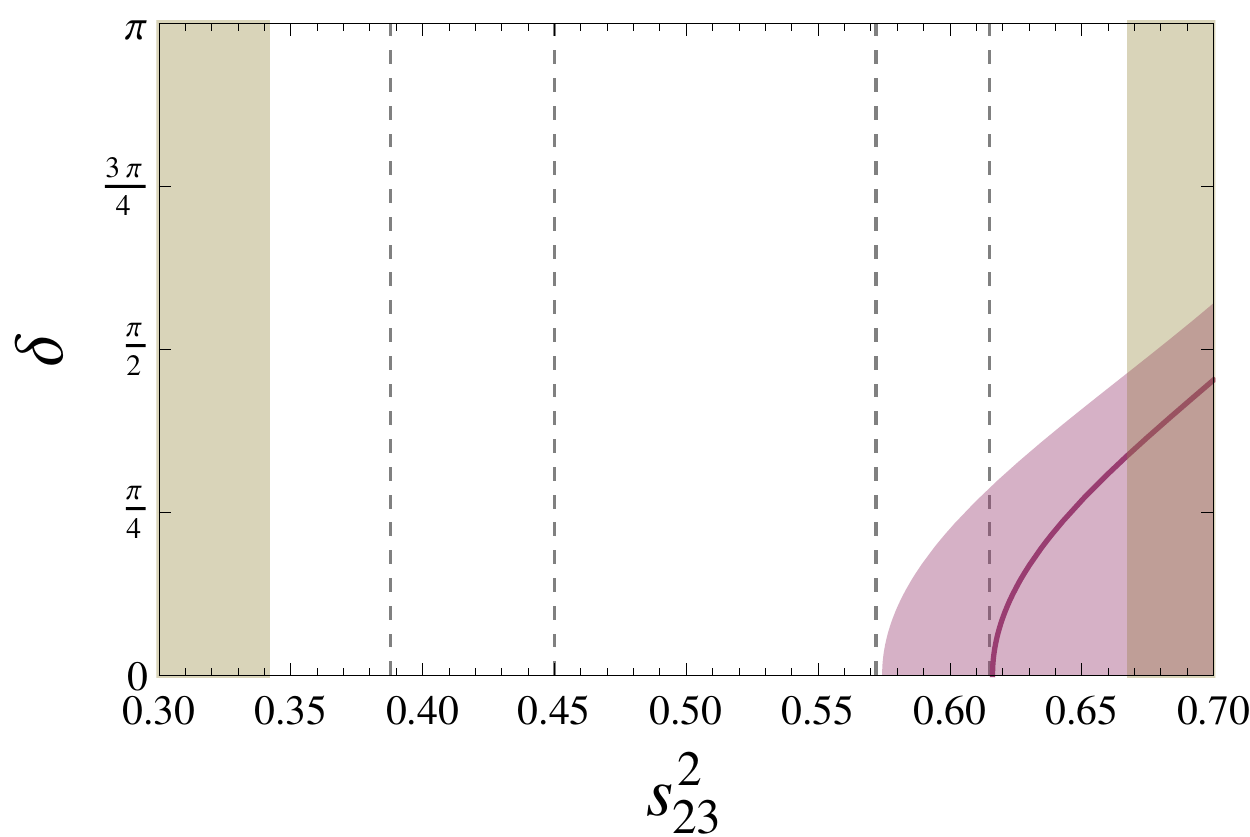}
\eec
\caption{\it 
The correlation between $\sin^2\theta_{23}$ and $\delta$ defined by \eq{pred12}. The thick purple line (the shaded purple region) corresponds
to the best fit values (the $3\sigma$ allowed ranges) for $\theta_{12}$ and $\theta_{13}$. The vertical yellow bands (dashed lines) limit the $3\sigma$
($1\sigma$) allowed range for $\sin^2\theta_{23}$.
}\label{last}
\end{figure}

By using \eq{pmns}, one can easily check that $\theta_{12}$, $\theta_{13}$, $\theta_{23}$ and $\delta$ are a function of $\theta_\nu$, $\theta_e$ and a
unique combination of phases, $2\alpha-\beta$, while the two physical
Majorana-type phases are a function of the other independent combination of  phases too.
The correlation between the mixing angles and $\delta$ depends on the assignment of the neutrino masses: 
we discuss first the case $m_1 = |m_-|$, $m_2=|c|$ and $m_3=|m_+|$, that corresponds to normal mass ordering. 
Comparing the predicted form of $U_{PMNS}$ with its usual parameterization, we obtain 
\beq
2 s_{12}^2 c_{23}^2 s_{13}^2 + 4 s_{12} c_{12} s_{23} c_{23} s_{13}  \cos\delta + 2 c_{12}^2 s_{23}^2  = 1 ~.
\label{pred12}\eeq
This relation is illustrated in Fig.~\ref{last}.
For the best fit values for $\theta_{12}$ and $\theta_{13}$, 
one finds $\cos\delta\gtrsim 0.49$, 
and for $\delta=0$ one reaches the minimal value $\sin^2\theta_{23}\simeq 0.616$, on the edge of the $1\sigma$ preferred region in the second octant.

Adding also $\Delta m^2_{21}$ and $\Delta m^2_{31}$ as input parameters, one can determine the lightest neutrino mass $m_1$ and the 
$0\nu2\beta$-decay effective mass $m_{ee}$ as a function of $\delta$ (or equivalently $\theta_{23}$) and of the extra phase $\alpha$.
This is because, in this scenario, in the basis where $M_e$ is diagonal  $M_\nu$ depends on six parameters, therefore, after the four $p_a$'s are fixed,
one can still vary two independent CP-violating phases.
One can derive the explicit dependence of $m_1$ and $m_{ee}$ on $\alpha$, by generalizing our procedure of section \ref{preds}. 
For simplicity we limit ourselves to the CP-conserving case, because \eq{pred12} favours a value of  $\delta$  close to zero, and because
the value of $m_{ee}$ for a CP-violating value of $\alpha$ is intermediate between its two values in the CP-conserving case: 
for $\delta=0$ we obtain $m_1\simeq 0.036$ eV and, for $\alpha=0$ ($\alpha=\pi/2$), $m_{ee} \simeq 0.012$ eV
($m_{ee} \simeq 0.034$ eV).

We remark that \eq{pred12}
follows from our minimality assumptions, that is to say, it is the consequence of the simplest realization of a 1+2 flavour symmetry.
This implies a non-maximal value of $\theta_{23}$ in the second octant.
It is interesting to compare \eq{pred12} with the relations among the mixing angles $\theta_{ij}$'s and the phase $\delta$, that have been 
obtained in \Ref{Hernandez:2012ra}. In that approach one assumes that the flavour symmetry group is generated by
a $Z_2$ symmetry in the neutrino sector plus a $Z_m$ symmetry in the charged lepton sector, with $m\ge 3$, and this leads to two constraints on the values
of  $\theta_{ij}$'s and $\delta$. The same analysis can be carried on in the present case, with $m=2$, and we find that one of the two constraints is trivially satisfied,
while the other is given by \eq{pred12}.

If the assignment of the neutrino mass eigenvalues is $m_1=|m_-|$, $m_2=|m_+|$ and $m_3=|c|$, both normal and inverse orderings are possible, 
and \eq{pred12} is replaced by 
\beq
2c_{13}^2c_{23}^2=1~.
\eeq
In this case the best fit value of $\theta_{13}$ corresponds to an almost maximal $2-3$ mixing, with $\sin^2\theta_{23} \simeq 0.49$,
a value (slightly) disfavoured by present data.
The other assignments of the three neutrino mass eigenvalues are not viable.

\section{Conclusions}

We attempted to provide a modern view on the lepton mixing, and to propose some keys to interpret it. 
The underlying flavour dynamics may be intricate and difficult to disentangle, especially if it occurs at very large energy scales. 
Nonetheless, the effective structure of the lepton mass matrices may well be simple and depend on a small number of parameters.
As a matter of fact, this must be the case for a flavour model to be testable.

Thus, it is useful to identify the minimal lepton mass matrices $M_\nu$ and $M_e$ compatible with those observables that have been measured precisely:
the charged lepton masses, the mixing angles $\theta_{12}$ and $\theta_{13}$, and the neutrino mass squared differences $\Delta m^2_{ij}$. 
Such viable flavour structures imply definite predictions for the neutrino mass observables that are not precisely measured yet: the mixing angle $\theta_{23}$,
the CP-violating phase $\delta$, the lightest neutrino mass $m_\text{light}$ and the $0\nu2\beta$-decay effective mass $m_{ee}$.
We proposed a general procedure to illustrate the correlations between these observables.

\begin{figure}[t]
\bec\begin{tabular}{r}
\includegraphics[width=0.6\linewidth]{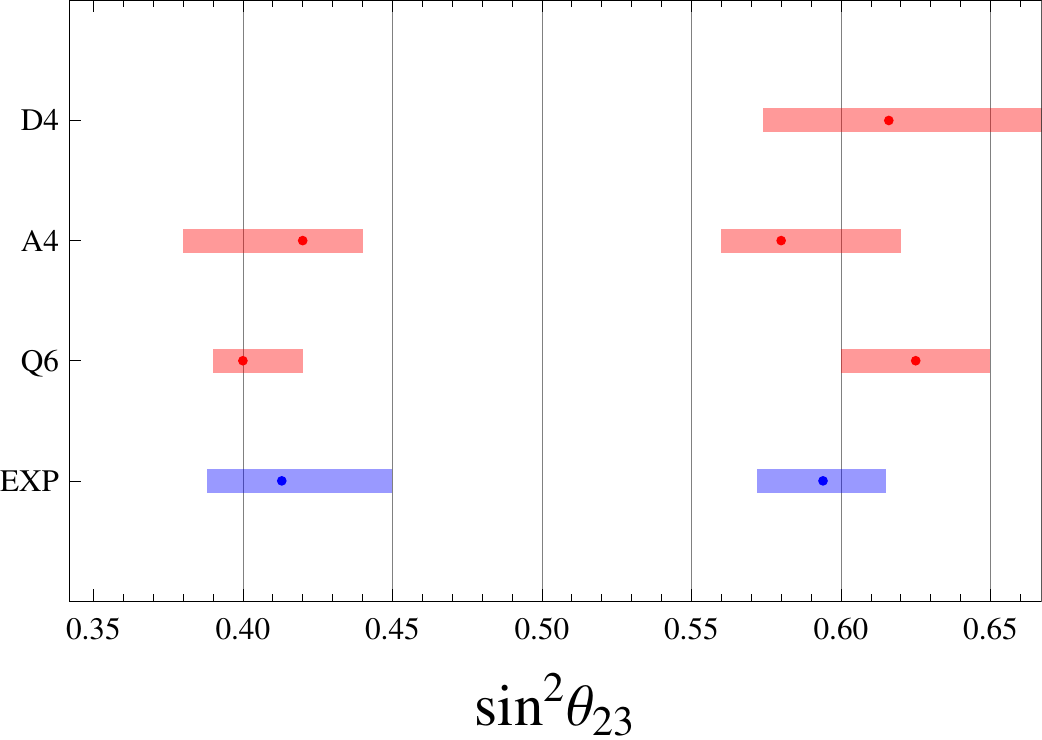}
\end{tabular}\eec
\caption{\it
The interval for $\sin^2\theta_{23}$ displayed here is the 
$3\sigma$ allowed range from a global fit of oscillation experiments  \cite{GonzalezGarcia:2012sz}. 
The lowest, blue dots (bands) are the central values (the $1\sigma$ ranges) in the two octants.
The upper, red dots (bands) are the predictions of our models, assuming the central values (the $3\sigma$ ranges) for the four parameters 
in \eq{pa}.
The three predictions correspond to
\eq{matrixA} ($Q_6$ model),   \eq{eq:zee} ($A_4$ model),
and \eq{pred12} with $\delta=0$ ($D_4$ model with no CP violation).
}\label{predictions}\end{figure}

In particular, it is intriguing that several minimal flavour structures imply a deviation from maximal $2-3$ mixing of the size that is presently suggested by the data.
We identified a few classes of mass matrices with this property:
\begin{itemize}
\item A matrix $M_\nu$ with two zero entries and two other entries equal to each other, in the basis where $M_e$ is diagonal.
There are two remarkable cases:
[i] $M_{\mu\mu}=M_{\tau\tau}=0$ and $M_{e\mu(e\tau)}=M_{\mu\tau}$ or $M_{e\mu(e\tau)}=M_{ee}$ (section \ref{preds});
[ii] $M_{ee}=M_{e\mu(e\tau)}=0$ and $M_{\mu\tau}=M_{\mu\mu(\tau\tau)}$ (appendix).
We pointed out the properties of the flavour symmetry group that is needed to explain these matrix structures (section \ref{sec:dp3}). 
\item An off-diagonal contribution to $M_\nu$ plus a contribution proportional to the identity, in the basis where $M_e$ is diagonal.
Such structure is tailor-made for an $A_4$ flavour symmetry (section \ref{sec:zeeplusd}).
\item Matrices $M_\nu$ and $M_e$ that respect each a different $Z_2$ symmetry, in such a way that only one free mixing angle
is present in each sector. The minimal realization is provided by a $D_4$ symmetry, with the three lepton families transforming in the $1+2$
irreducible representations (section \ref{1+2}). 
\end{itemize}
The three specific models that we presented predict $\theta_{23}$ close to its best fit values, as summarized in Fig.~\ref{predictions}.
Thus, the future precision measurements of this mixing angle  will be able to confirm or rule out these scenarios.
In order to discriminate among them, one shall use the correlated predictions for the mass ordering as well as for $\delta$, $m_\text{light}$
and $m_{ee}$, that could be all precisely measured within the next decade.

\section*{Acknowledgements}

We thank Sacha Davidson, Ernest Ma, Thomas Schwetz, Damien Tant and Mariam Tortola for providing us with motivations, helpful criticism and interpretation of the data.
AVM acknowledges partial support from the IN2P3 project  ``Th\'eorie LHC France''.
MF thanks the Galileo Galilei Institute (INFN, Florence), the FP7 European ITN project ``Invisibles" (PITN-GA-2011-289442-INVISIBLES)
and the Theoretical Physics Group (CERN, Geneva)  for partial support and hospitality.

\appendix
\section*{Appendix}
\label{sec:2zero}

Here we detail the procedure to derive the functions $x_i(x_j,p_a)$, defined in section \ref{preds}, for 
a special class of Majorana neutrino mass matrices $M_\nu$ that depend on five physical parameters:
the $3\times 3$ symmetric matrices with two independent entries equal to zero.
There are 15 such matrix structures, and only 7 of them are compatible with oscillation data, as first shown 
in \Ref{Frampton:2002yf} (see \Ref{Fritzsch:2011qv} for a recent analysis).

The four constraints relating the observables (3 neutrino mass eigenvalues, 3 mixing angles and 3 phases) have the form
\beq
0 = M_{\alpha\beta} = \sum_i (U^*_{PMNS})_{\alpha i} (U^*_{PMNS})_{\beta i} m_i ~, 
\eeq
for two independent complex entries $M_{\alpha_1\beta_1}$ and $M_{\alpha_2\beta_2}$.
Since these equations are linear in $m_i$, they can be rewritten in the form
\beq
\frac{m_1}{m_3} e^{i\varphi_{13}} = f^*_{13}(\theta_{ij},\delta)~,~~~\frac{m_2}{m_3} e^{i\varphi_{23}} = f^*_{23}(\theta_{ij},\delta)~,
\eeq
where $\varphi_{ij}$ are the Majorana phases and
the functions $f_{ij}$ depend on the choice of $(\alpha_1,\beta_1)$ and $(\alpha_2,\beta_2)$,
\begin{align}
f_{13} & = \frac{U_{\alpha_1 2}U_{\beta_1 2}U_{\alpha_2 3}U_{\beta_2 3} - U_{\alpha_1 3}U_{\beta_1 3}U_{\alpha_2 2}U_{\beta_2 2}}{U_{\alpha_1 1}U_{\beta_1 1}U_{\alpha_2 2}U_{\beta_2 2}-U_{\alpha_1 2}U_{\beta_1 2}U_{\alpha_2 1}U_{\beta_2 1}}~,\\
f_{23} & = \frac{U_{\alpha_1 3}U_{\beta_1 3}U_{\alpha_2 1}U_{\beta_2 1}-U_{\alpha_1 1}U_{\beta_1 1}U_{\alpha_2 3}U_{\beta_2 3}}{U_{\alpha_1 1}U_{\beta_1 1}U_{\alpha_2 2}U_{\beta_2 2}-U_{\alpha_1 2}U_{\beta_1 2}U_{\alpha_2 1}U_{\beta_2 1}}~, 
\end{align}
where $U_{\alpha i}$ are the $U_{PMNS}$ entries without the Majorana phases. 
The neutrino mass squared differences can be expressed in terms of the neutrino masses as
\begin{equation}\left\{\begin{aligned}
\Delta m_{21}^2 &= m_3^2 (|f_{23}|^2 - |f_{13}|^2),\\
\Delta m_{31}^2 &= m_3^2 (1 - |f_{13}|^2).
\end{aligned}\right.
\label{eq:Del}
\end{equation}
These equations 
are linear in $m_3^2$ and linear or quadratic in $\cos\delta$. Therefore the 
system has at most 2 solutions (for each given ordering of the mass spectrum) of the form 
\begin{equation}
m_3^2 = m_3^2 (\theta_{23}, p_a),\qquad
\cos\delta = \cos\delta (\theta_{23}, p_a),
\label{eq:m3-cd}
\end{equation}
where the input parameters $p_a$'s are defined in \eq{pa}. 
Each solution is physical only when it fulfills the requirements $m_3^2>0$ and $|\cos\delta|\le 1$. 
The effective $0\nu2\beta$-decay mass parameter can be written as 
\begin{equation}
m_{ee} = m_3 \left| U^{*2}_{e1} f^*_{13} + U^{*2}_{e2} f^*_{23} + U^{*2}_{e3} \right|
= m_{ee}(\theta_{23},p_a)~,
\end{equation}
where one has inserted the expressions for $m_3$ and $\delta$ given by \eq{eq:m3-cd}.
Analogously, any matrix element $M_{\alpha\beta}$ can be calculated as a function of $\theta_{23}$ and the $p_a$'s.
The functions $x_i(\theta_{23}, p_a)$ for the observables $x_i=m_3,\delta,m_{ee}$ 
provide a parametric representation for the functions $x_i(x_j,p_a)$, for any pair $i,j$.
This concludes the analytic computation of the observables for any $M_\nu$ with two zero entries.

\begin{figure}[t]
\bec\begin{tabular}{c}
\includegraphics[width=0.45\linewidth]{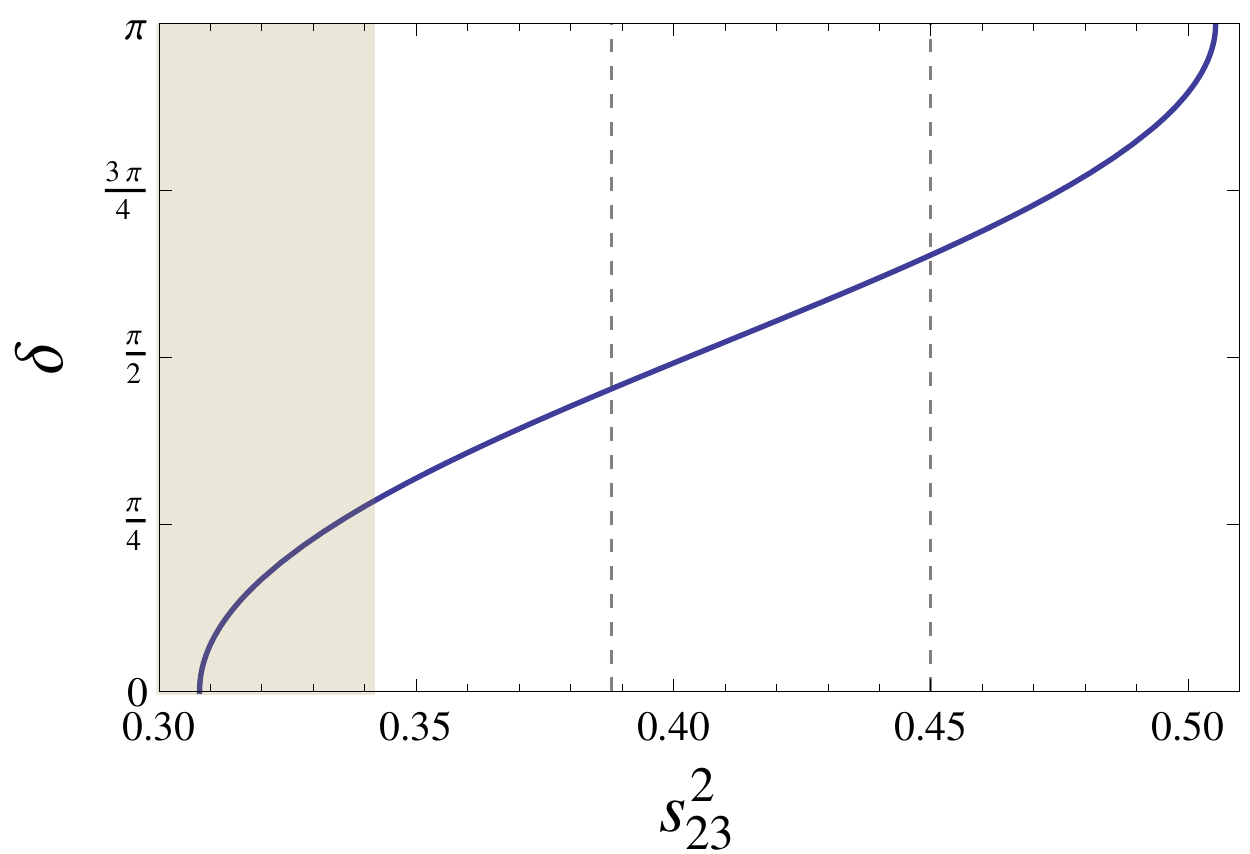}\quad
\includegraphics[width=0.45\linewidth]{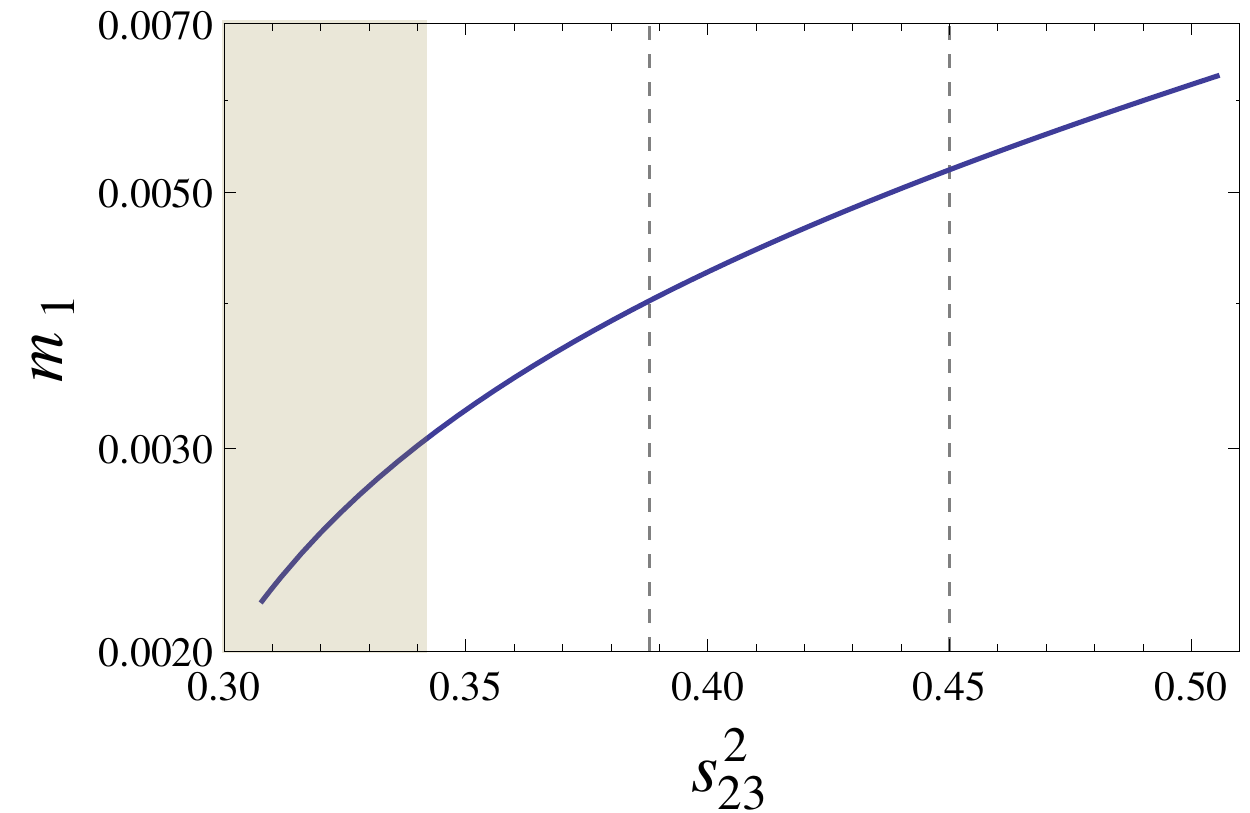}\\
\includegraphics[width=0.45\linewidth]{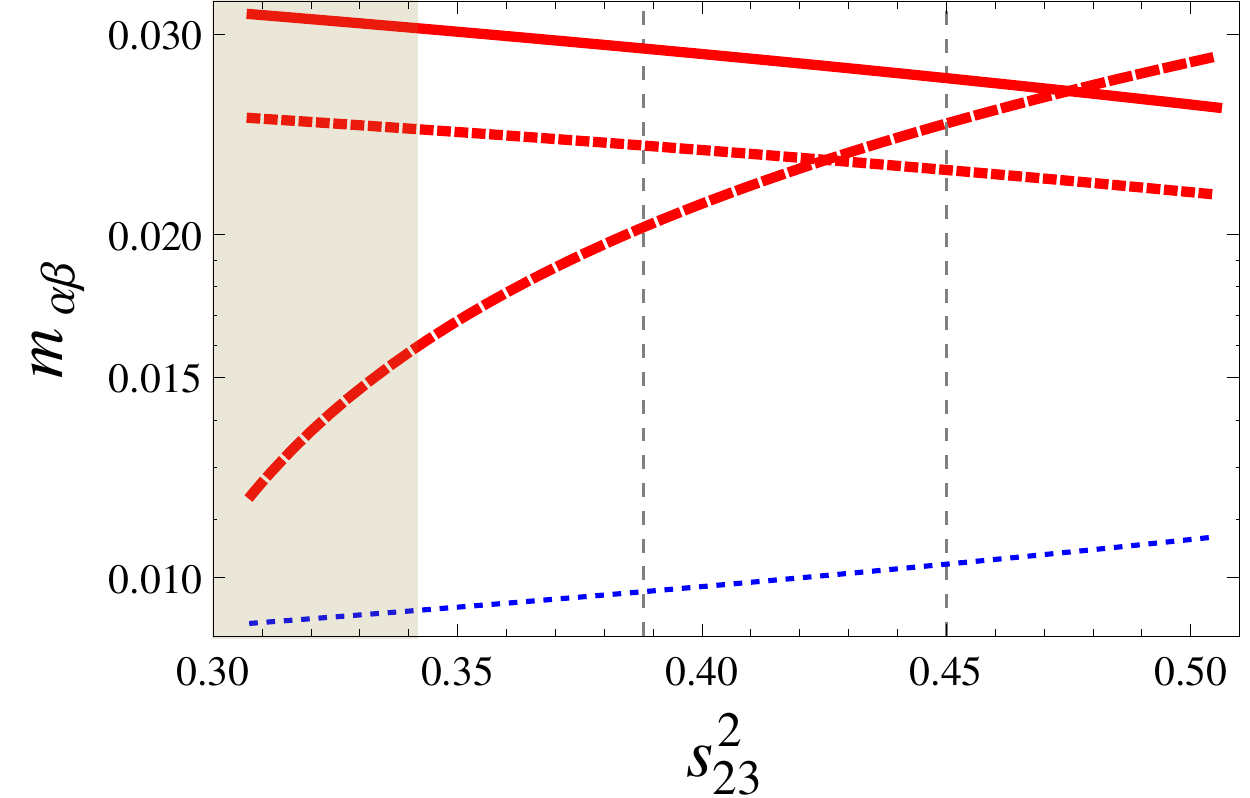}
\end{tabular}\eec
\caption{\it
The CP-violating Dirac phase $\delta$ (top left), the lightest neutrino mass $m_1$ in eV (top right) and the 
absolute values of the neutrino mass matrix elements $m_{\alpha\beta}$ in eV
(bottom), as a function of $\sin^2\theta_{23}$, in the case $m_{ee}=m_{e\mu}=0$. The ordering of the mass spectrum is necessarily normal. 
The yellow bands are excluded by the $3\sigma$ bound on $\theta_{23}$; the vertical dashed lines limit the $1\sigma$ preferred region for $\theta_{23}$.
In the bottom panel, the thin dotted blue line corresponds to $m_{e\tau}$, the thick dotted red line to $m_{\mu\tau}$, 
the thick dashed red line to $m_{\mu\mu}$ and the thick solid red line to $m_{\tau\tau}$.}
\label{figA}\end{figure}

The results for $M_{\mu\mu}=M_{\tau\tau}=0$ (case $C$) were presented in section \ref{preds}. Here we present the results for the 6 other viable cases:
\bea
& A_1:~ M_{e\mu}=M_{ee}=0,\quad & A_2:~ M_{e\tau}=M_{ee}=0, \\
& B_1:~ M_{e\tau}=M_{\mu\mu}=0, \quad & B_2:~ M_{e\mu}=M_{\tau\tau}=0,\\
& B_3:~ M_{e\mu}=M_{\mu\mu}=0,\quad & B_4:~ M_{e\tau}=M_{\tau\tau}=0.
\eea
The matrices in the second column are obtained from those in the first column, by the exchange of the 2nd and 3rd rows and columns,
therefore their predictions are simply related by \eq{mutau}.
Thus, we will illustrate the results for the cases $A_1$, $B_1$ and $B_3$ only, that for
normal mass ordering correspond to $\theta_{23}$ in the first octant, as preferred by global fits \cite{Tortola:2012te,GonzalezGarcia:2012sz}.
 When inferring the results
for the `specular' cases, one should recall that the allowed experimental range
for $\theta_{23}$ is slightly asymmetric between the first and the second octant.

For simplicity we fix the input parameters  $p_a$'s to their best fit values:
the predictions are slightly different when the small uncertainties on these parameters are taken into account.
We present the plots for $m_\text{light}(\sin^2\theta_{23},p_a)$, $\delta(\sin^2\theta_{23},p_a)$ and $m_{\alpha\beta}(\sin^2\theta_{23},p_a)$, 
that are sufficient to infer
the correlation between any pair of observables $x_i$'s.

\begin{figure}[tb]
\begin{tabular}{cc}
\includegraphics[width=0.45\linewidth]{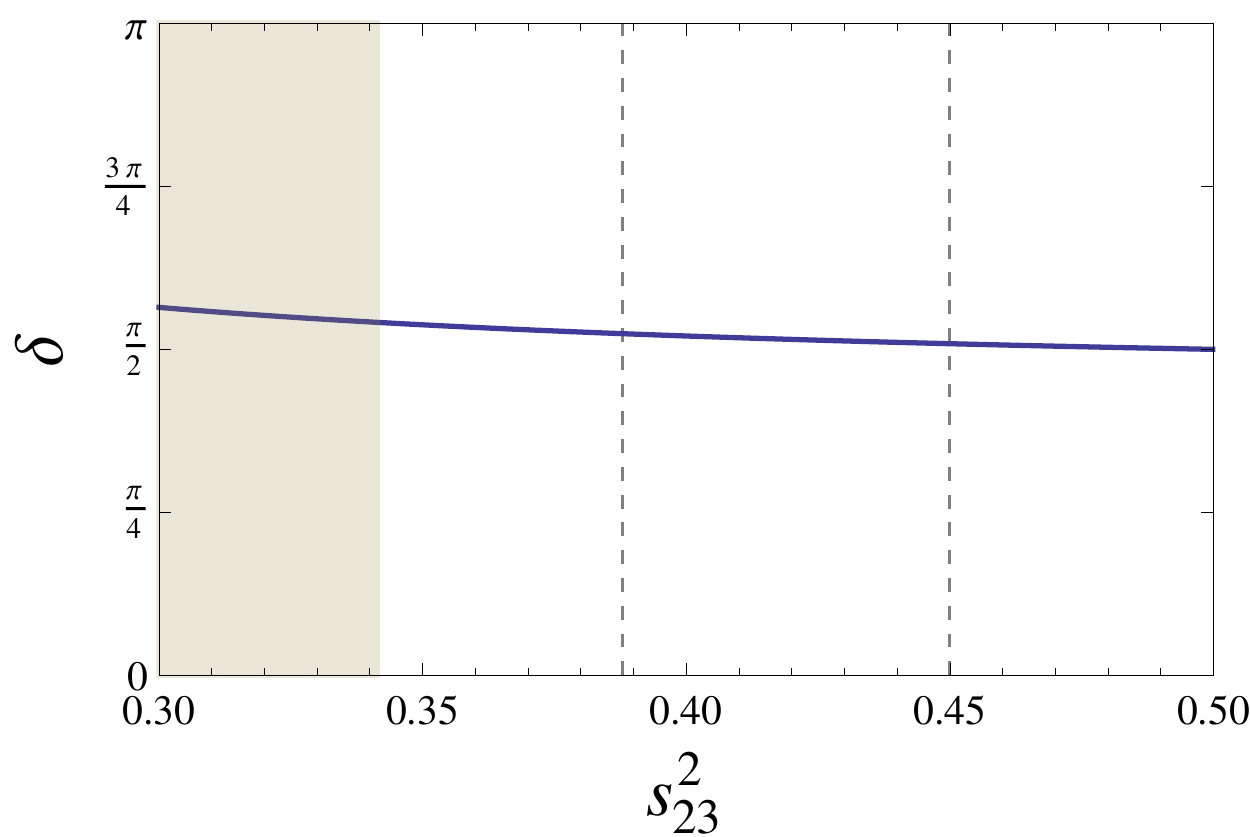}&
\includegraphics[width=0.45\linewidth]{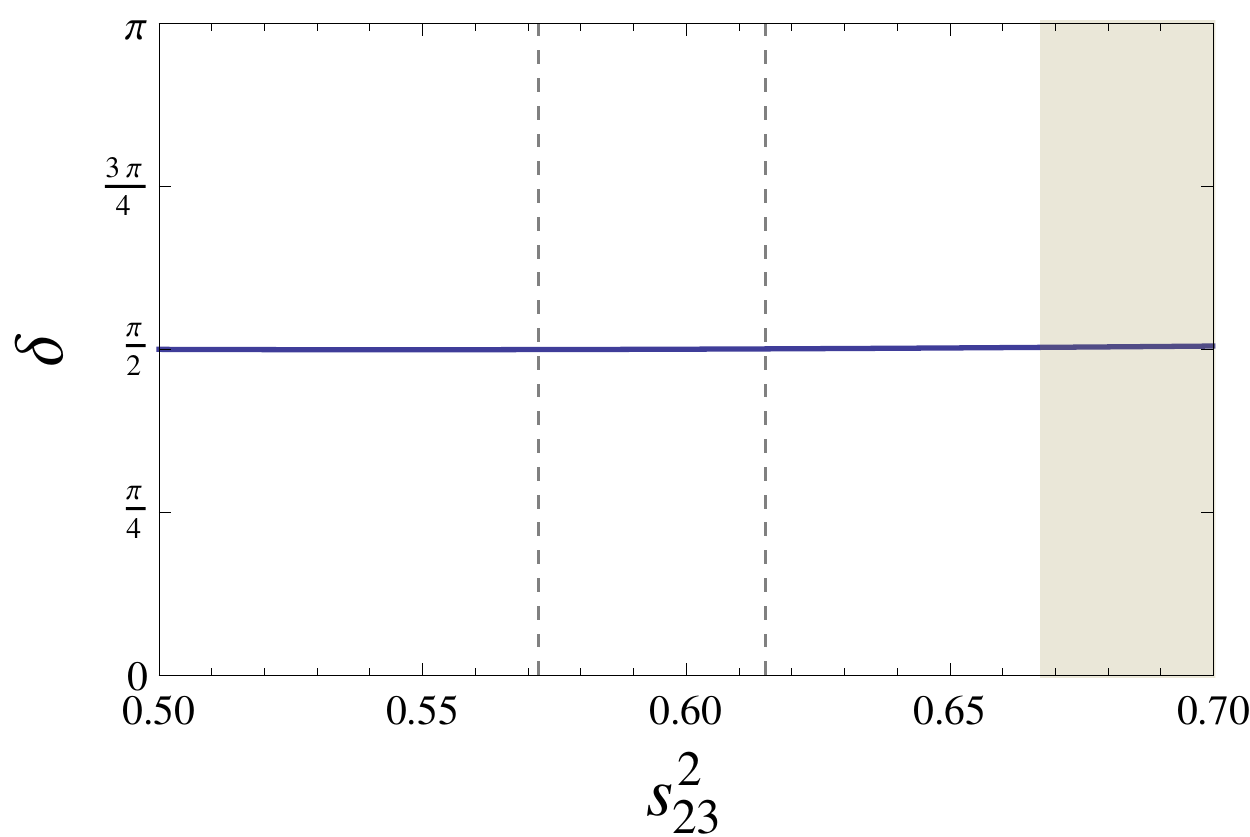}\\
\includegraphics[width=0.45\linewidth]{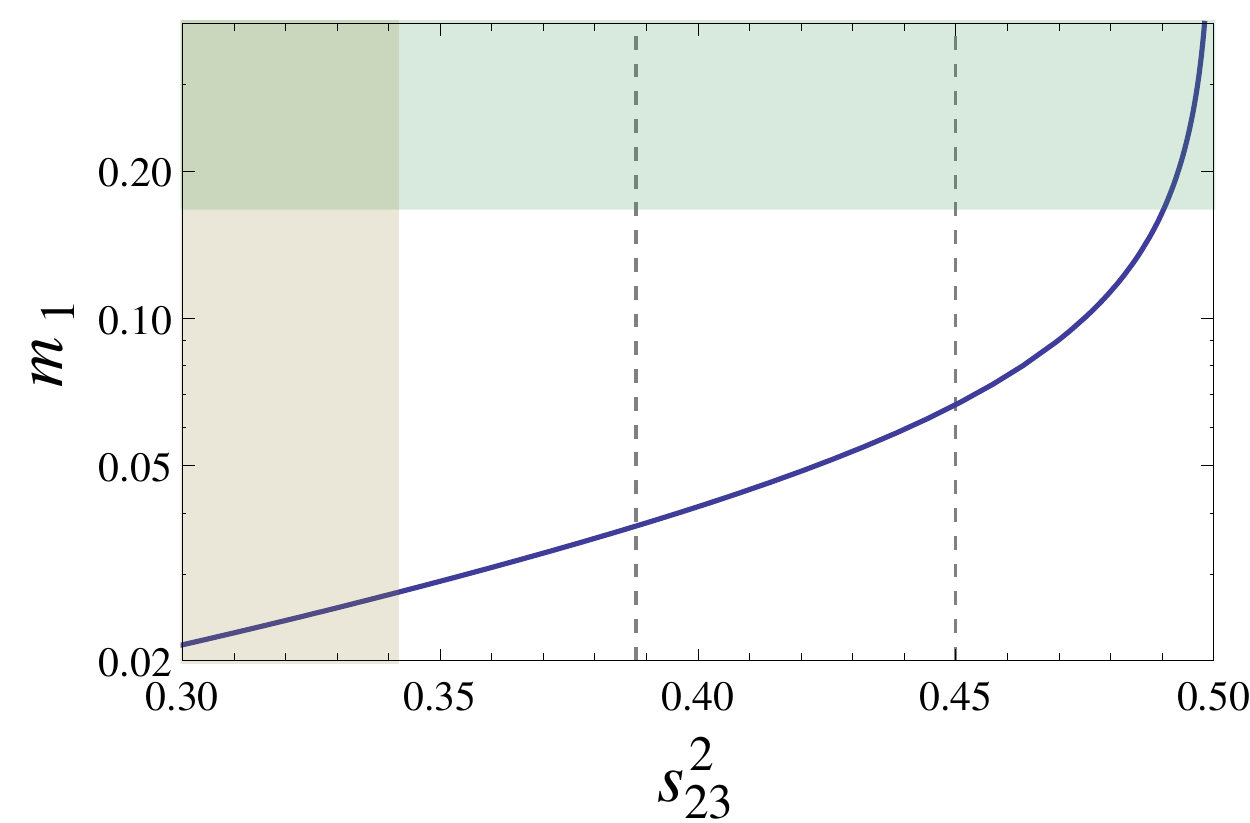}&
\includegraphics[width=0.45\linewidth]{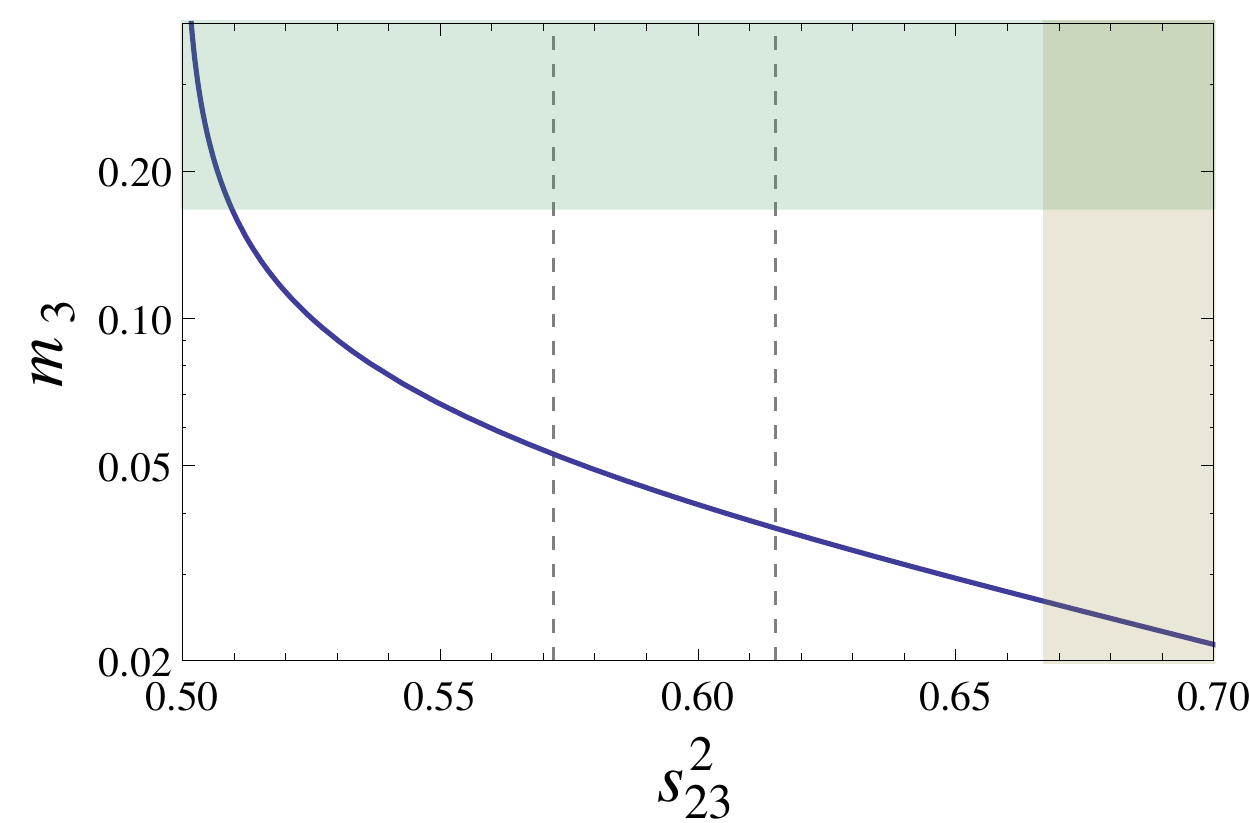}\\
\includegraphics[width=0.45\linewidth]{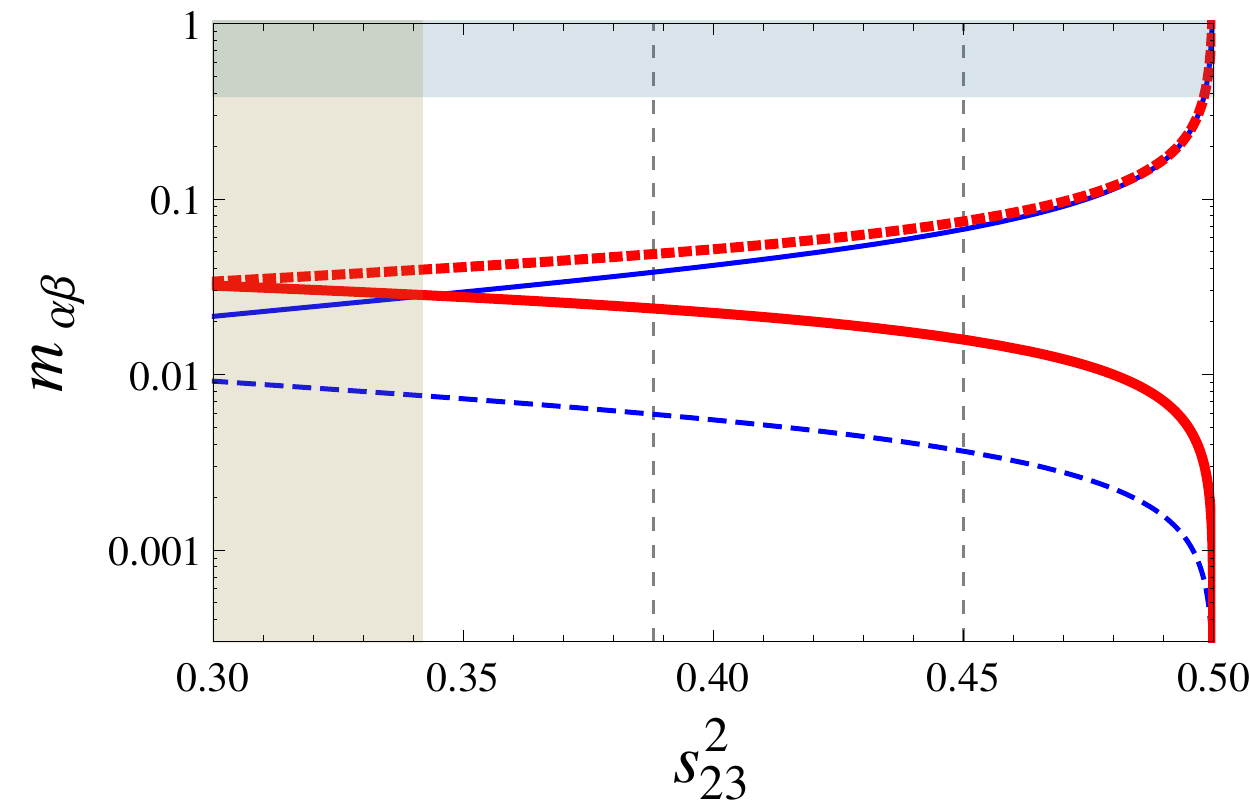}&
\includegraphics[width=0.45\linewidth]{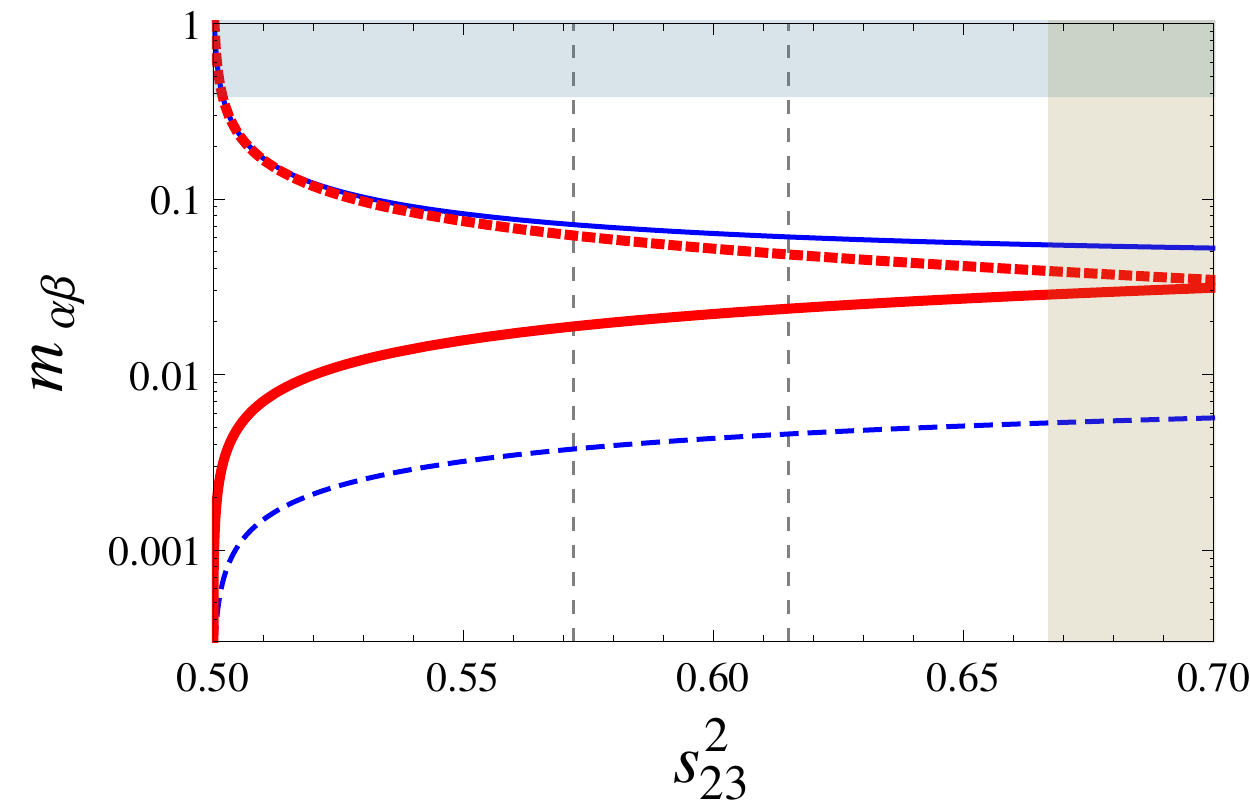}
\end{tabular}
\caption{\it
The same as in Fig.~\ref{figA}, but for the case $m_{\mu\mu}=m_{e\tau}=0$. The left-hand (right-hand) plots correspond to normal (inverse)
ordering. 
The green (blue) bands are excluded by the cosmological upper bound on $\sum m_i$ (the experimental upper bound on $m_{ee}$). 
In the bottom panels, the thin solid blue line corresponds to $m_{ee}$, the thin dashed blue line to $m_{e\mu}$, the thick dotted red line to $m_{\mu\tau}$ and
the thick solid red line to $m_{\tau\tau}$. }
\label{figB}\end{figure}

\begin{figure}[tb]
\begin{tabular}{cc}
\includegraphics[width=0.45\linewidth]{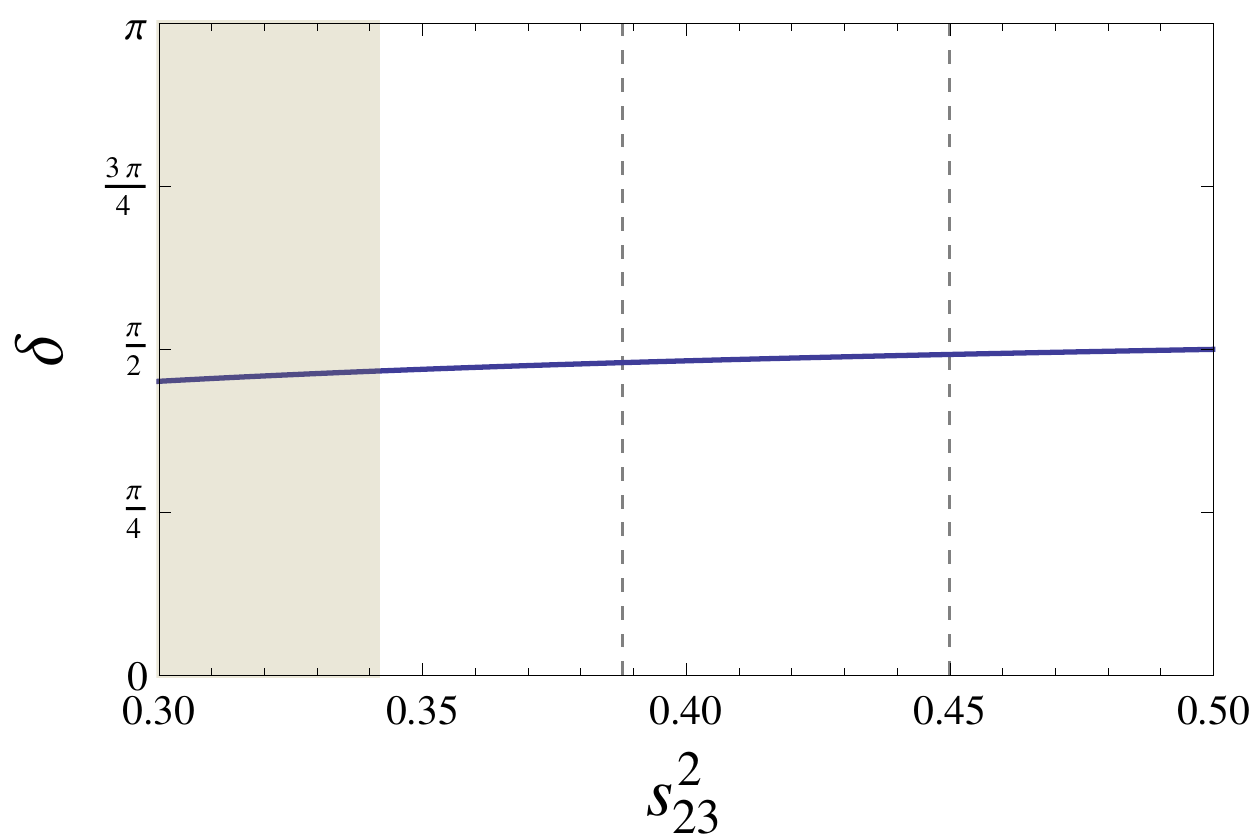}&
\includegraphics[width=0.45\linewidth]{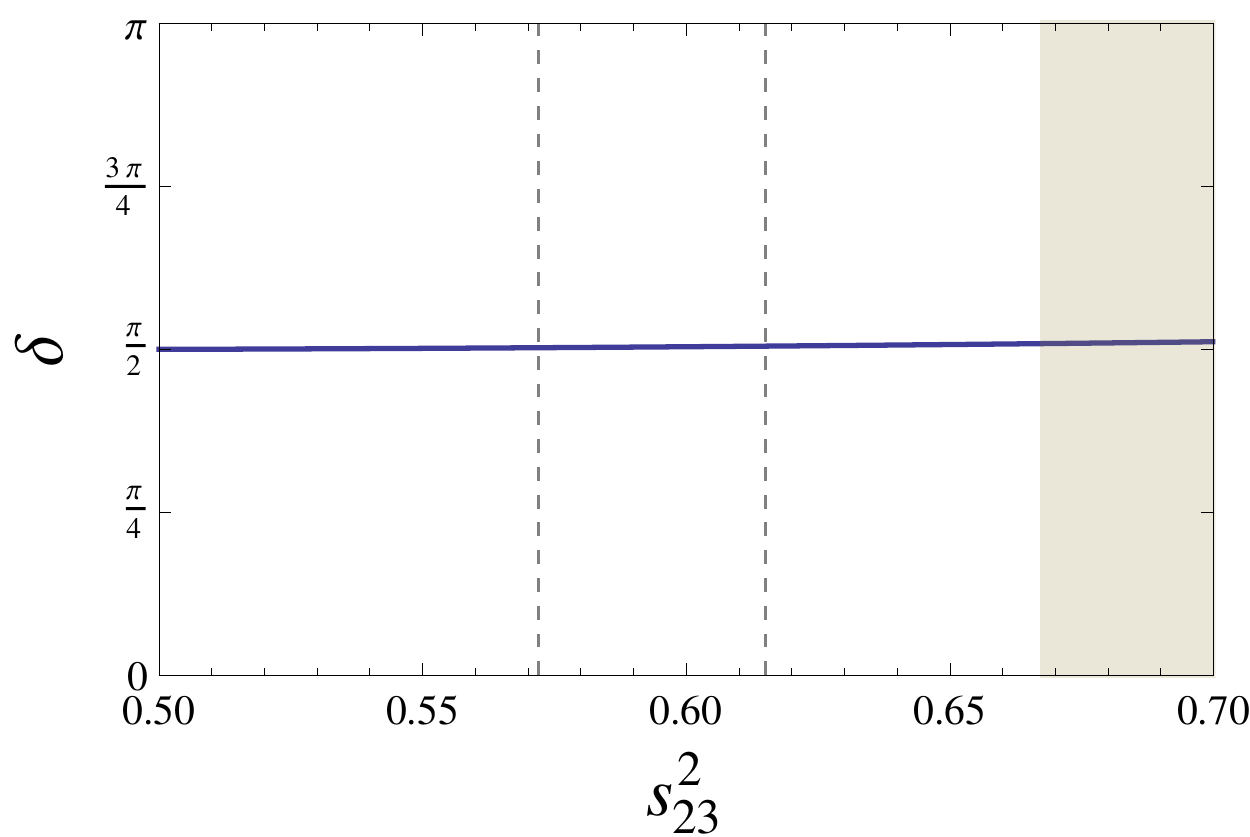}\\
\includegraphics[width=0.45\linewidth]{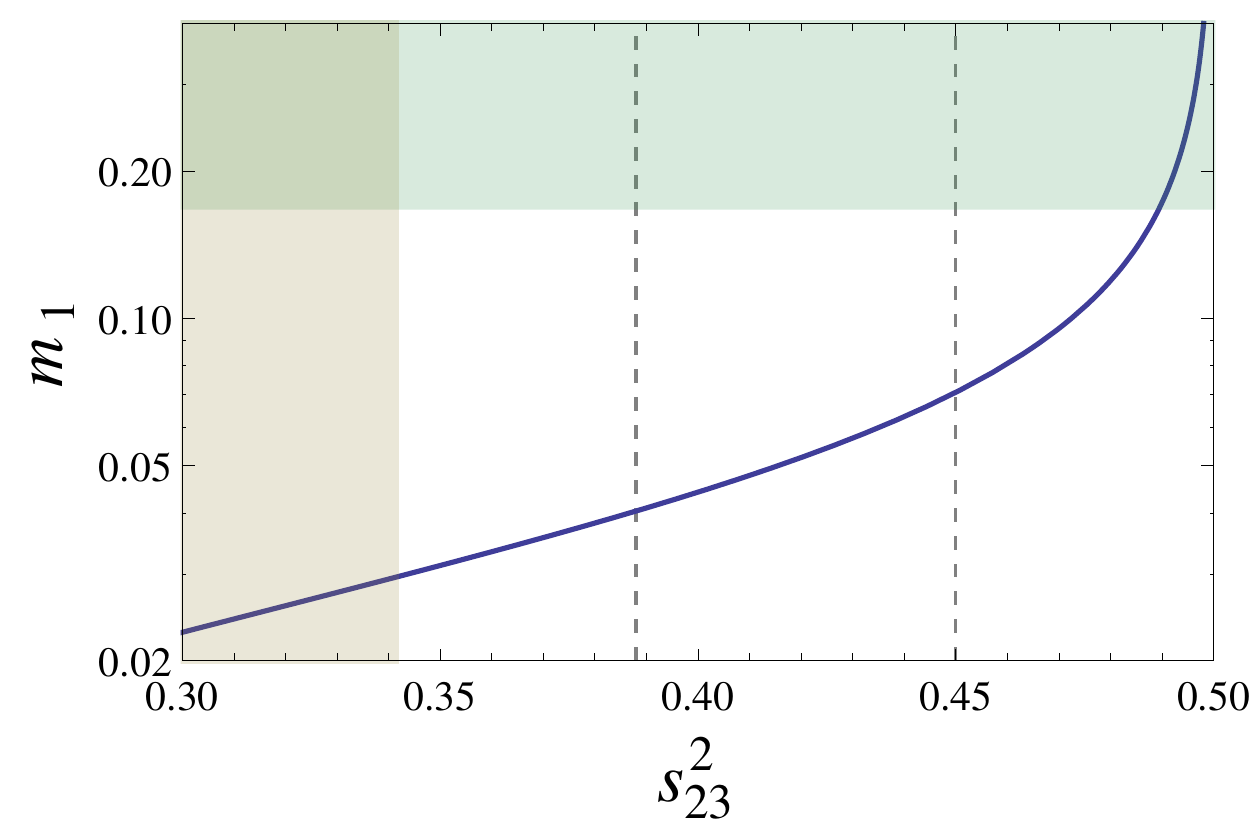}&
\includegraphics[width=0.45\linewidth]{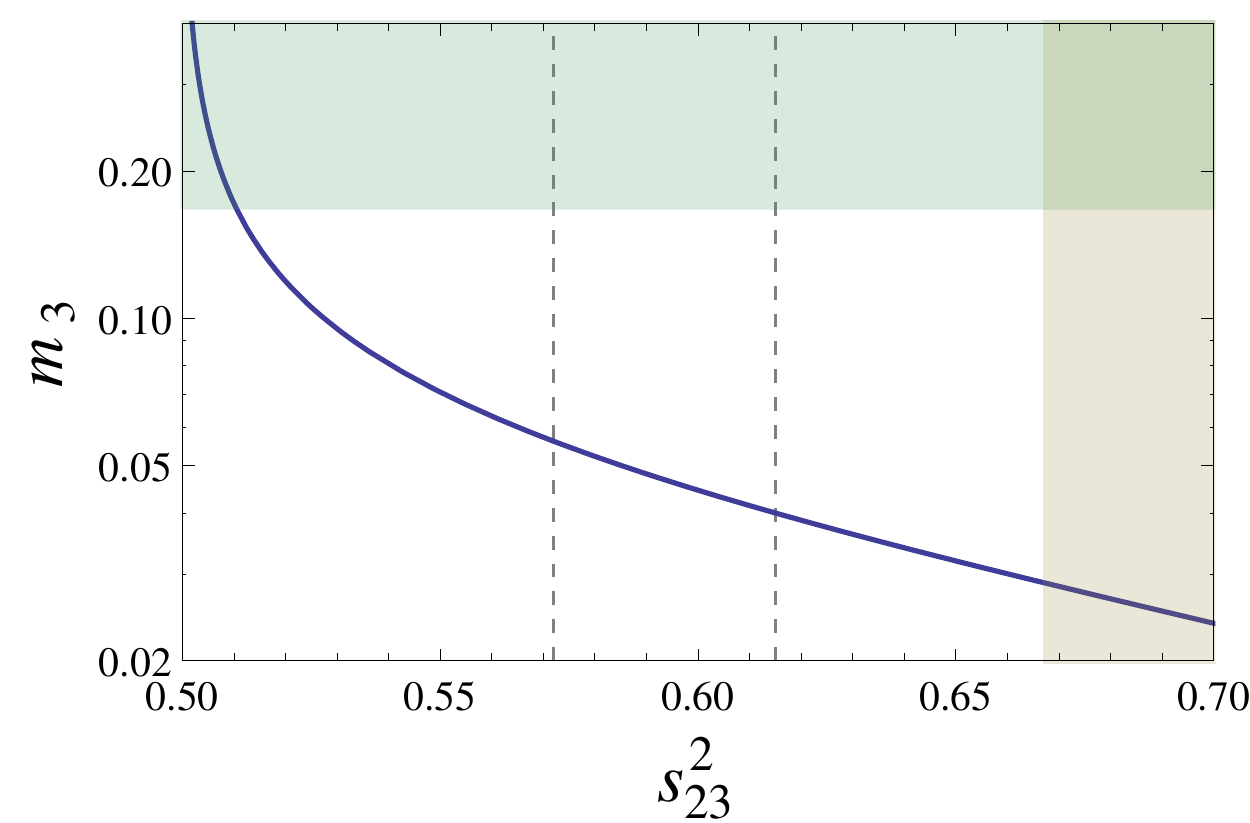}\\
\includegraphics[width=0.45\linewidth]{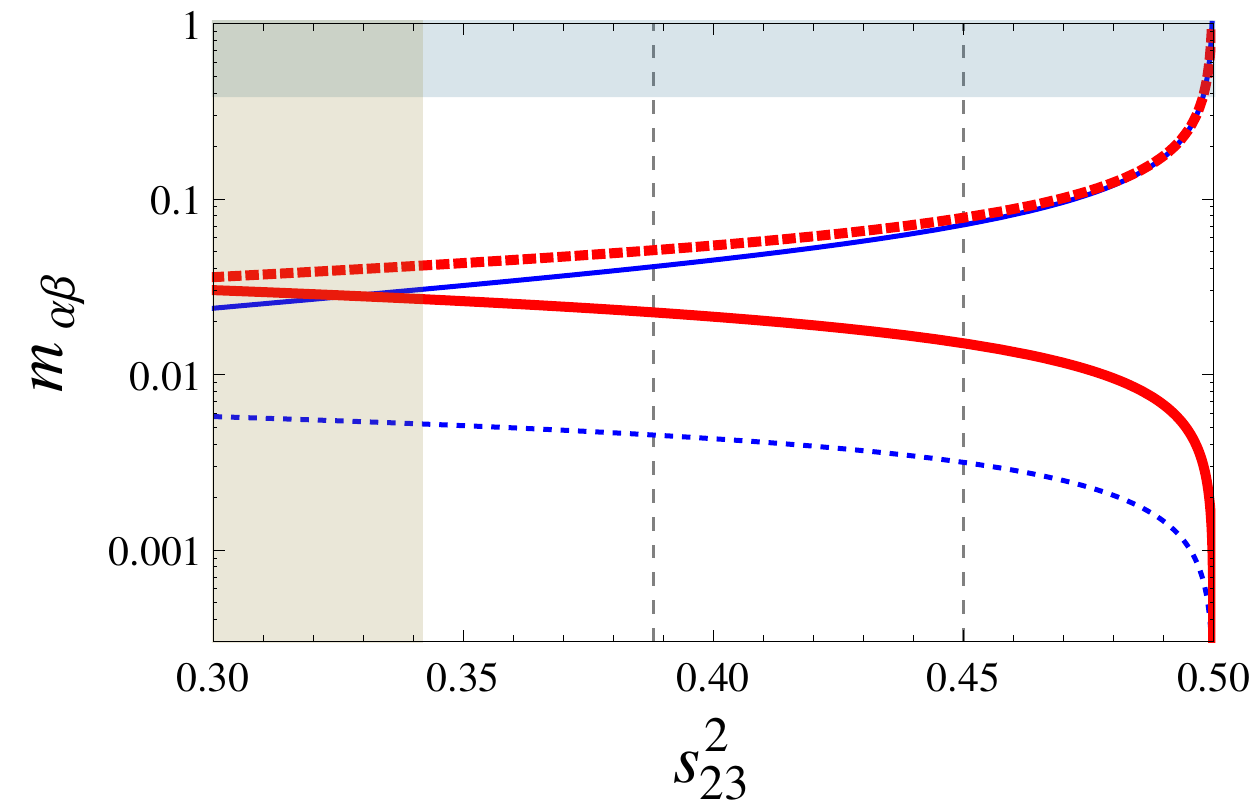}&
\includegraphics[width=0.45\linewidth]{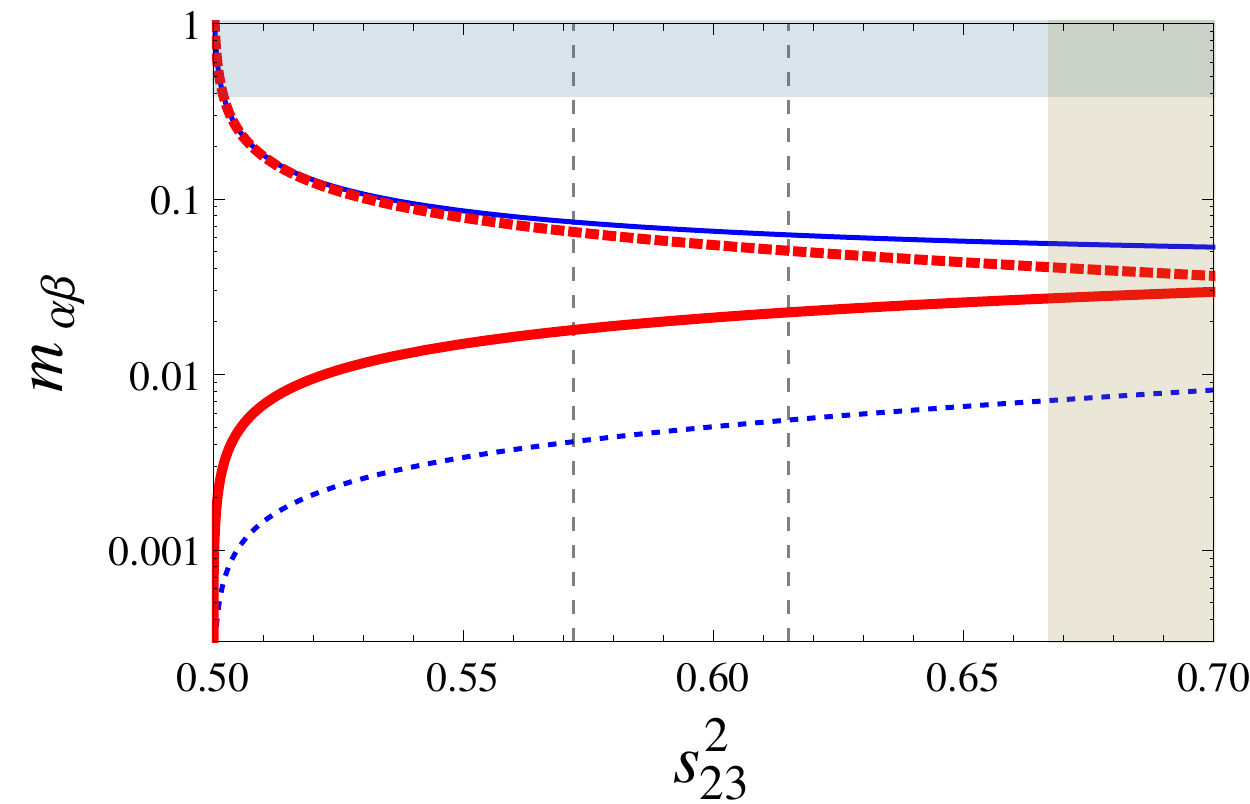}
\end{tabular}
\caption{\it
The same as in Figs.~\ref{figA} and \ref{figB}, but for the case $m_{\mu\mu}=m_{e\mu}=0$. 
In the bottom panels, the thin solid blue line corresponds to $m_{ee}$, the thin dotted blue line to $m_{e\tau}$, the thick dotted red line to $m_{\mu\tau}$ and
the thick solid red line to $m_{\tau\tau}$.}
\label{figC}\end{figure}
\begin{itemize}
\item The results for the case $A_1$ are illustrated in Fig.~\ref{figA}. 
\begin{itemize}
\item Only a normal ordering of the neutrino mass spectrum is allowed. 
\item The $2-3$ mixing satisfies $\sin^2\theta_{23} < 0.505$, however this upper bounds relaxes substantially when allowing the 
$p_a$'s to vary in their $3\sigma$ allowed range. 
\item The phase $\delta$ satisfies $\cos\delta < 0.62$. 
\item The lightest mass $m_1$ lies in the range $[3.1,\,6.3]\times 10^{-3}$ eV.
\item There is no contribution to the $0\nu2\beta$-decay rate.
\item In the $1\sigma$ preferred region for $\theta_{23}$, the equality $m_{\mu\mu}=m_{\mu\tau}$ can be realized. Therefore, as in the case $C$
discussed in section \ref{preds}, the vanishing of two matrix entries and the equality of two others imply the desired prediction for $\theta_{23}$.
One could build a model on the lines of the one in section \ref{sec:dp3}, in order to justify this matrix structure by a discrete family symmetry.
\end{itemize}
\item The results for the case $B_1$ are illustrated in Fig.~\ref{figB}. 
\begin{itemize}
\item Both normal and inverse ordering are allowed.
\item The angle $\theta_{23}$ lies in the first (second) in the case of normal (inverse) ordering. 
This statement remains valid when allowing the $p_a$'s to vary in their $3\sigma$ allowed range. 
\item The phase $\delta$ is very close to $\pi/2$ (maximal CP-violation), 
with $|\cos\delta| < 0.13$ (a similar bound holds when the $3\sigma$ uncertainties on the $p_a$'s are taken into account). 
\item There is a lower bound on the lightest neutrino mass,  $m_1 > 2.8 \times 10^{-2}$ eV for normal and $m_3 > 2.6 \times 10^{-2}$ eV for 
inverse ordering. 
\item There is a lower bound on the $0\nu2\beta$-decay effective mass, 
$m_{ee} > 2.8 \times 10^{-2}$ eV for normal or $m_{ee} > 5.5 \times 10^{-2}$ eV for inverse ordering. 
\end{itemize}
\item  The results for the case $B_3$ are illustrated in Fig.~\ref{figC}. 
They are qualitatively the same as in the case $B_1$, with minor modifications of the predicted range for the $x_i$'s.
\end{itemize}
In summary, our procedure allows to illustrate clearly all the predictions that follow from the assumption of two zero entries in $M_\nu$.
We confirm that 8 matrix structures of this type are already excluded by present data. The 7 allowed matrix structures, analyzed above, could be
ruled out by future measurements of the observables $x_i$'s. No matrix with three zero entries is allowed, because there is no intersection between
the predictions of the different matrices with two zero entries.



\begin{thebibliography}{99}

\bibitem{An:2012eh}
  F.~P.~An {\it et al.}  [DAYA-BAY Collaboration],
  Phys.\ Rev.\ Lett.\  {\bf 108} (2012) 171803
  [arXiv:1203.1669 [hep-ex]].
  J.~K.~Ahn {\it et al.}  [RENO Collaboration],
  Phys.\ Rev.\ Lett.\  {\bf 108} (2012) 191802
  [arXiv:1204.0626 [hep-ex]].
  Y.~Abe {\it et al.}  [DOUBLE-CHOOZ Collaboration],
  Phys.\ Rev.\ Lett.\  {\bf 108} (2012) 131801
  [arXiv:1112.6353 [hep-ex]].

\bibitem{Tortola:2012te}
  D.~V.~Forero, M.~Tortola and J.~W.~F.~Valle,
  Phys.\ Rev.\ D {\bf 86} (2012) 073012
  [arXiv:1205.4018 [hep-ph]].
  G.~L.~Fogli, E.~Lisi, A.~Marrone, D.~Montanino, A.~Palazzo and A.~M.~Rotunno,
  Phys.\ Rev.\ D {\bf 86} (2012) 013012
  [arXiv:1205.5254 [hep-ph]].

\bibitem{GonzalezGarcia:2012sz} 
  M.~C.~Gonzalez-Garcia, M.~Maltoni, J.~Salvado and T.~Schwetz,
  JHEP {\bf 1212}, 123 (2012)
  [arXiv:1209.3023 [hep-ph]].


\bibitem{Altarelli:2010gt}
  G.~Altarelli and F.~Feruglio,
  Rev.\ Mod.\ Phys.\  {\bf 82} (2010) 2701
  [arXiv:1002.0211 [hep-ph]].


\bibitem{Ma:2011yi}
  E.~Ma and D.~Wegman,
  Phys.\ Rev.\ Lett.\  {\bf 107} (2011) 061803
  [arXiv:1106.4269 [hep-ph]].
  R.~d.~A.~Toorop, F.~Feruglio and C.~Hagedorn,
  Phys.\ Lett.\ B {\bf 703} (2011) 447
  [arXiv:1107.3486 [hep-ph]].
  S.~-F.~Ge, D.~A.~Dicus and W.~W.~Repko,
  Phys.\ Rev.\ Lett.\  {\bf 108} (2012) 041801
  [arXiv:1108.0964 [hep-ph]].
  D.~A.~Eby and P.~H.~Frampton,
  Phys.\ Rev.\ D {\bf 86} (2012) 117304
  [arXiv:1112.2675 [hep-ph]].
  I.~de Medeiros Varzielas and G.~G.~Ross,
  JHEP {\bf 1212} (2012) 041
  [arXiv:1203.6636 [hep-ph]].
  S.~M.~Boucenna, S.~Morisi, M.~Tortola and J.~W.~F.~Valle,
  Phys.\ Rev.\ D {\bf 86} (2012) 051301
  [arXiv:1206.2555 [hep-ph]].
  G.~Altarelli, F.~Feruglio, I.~Masina and L.~Merlo,
  JHEP {\bf 1211} (2012) 139
  [arXiv:1207.0587 [hep-ph]].
  S.~Bhattacharya, E.~Ma, A.~Natale and D.~Wegman,
  Phys.\ Rev.\ D {\bf 87} (2013) 013006
  [arXiv:1210.6936 [hep-ph]].
  F.~Feruglio, C.~Hagedorn and R.~Ziegler,
  arXiv:1211.5560 [hep-ph].

\bibitem{Hernandez:2012ra}
  D.~Hernandez and A.~Y.~.Smirnov,
  Phys.\ Rev.\ D {\bf 86} (2012) 053014
  [arXiv:1204.0445 [hep-ph]].
  D.~Hernandez and A.~Y.~.Smirnov,
  arXiv:1212.2149 [hep-ph].

\bibitem{Barger:2001yr}
  V.~Barger, D.~Marfatia and K.~Whisnant,
  Phys.\ Rev.\ D {\bf 65} (2002) 073023
  [hep-ph/0112119].
%
  P.~Huber, M.~Maltoni and T.~Schwetz,
  Phys.\ Rev.\ D {\bf 71} (2005) 053006
  [hep-ph/0501037].
%
  M.~Ishitsuka, T.~Kajita, H.~Minakata and H.~Nunokawa,
  Phys.\ Rev.\ D {\bf 72} (2005) 033003
  [hep-ph/0504026].
  P.~Coloma, P.~Huber, J.~Kopp and W.~Winter,
  arXiv:1209.5973 [hep-ph].


\bibitem{Lesgourgues:2012uu}
  J.~Lesgourgues and S.~Pastor,
  Adv.\ High Energy Phys.\  {\bf 2012} (2012) 608515
  [arXiv:1212.6154 [hep-ph]].
  Z.~Hou, C.~L.~Reichardt, K.~T.~Story, B.~Follin, R.~Keisler, K.~A.~Aird, B.~A.~Benson and L.~E.~Bleem {\it et al.},
  arXiv:1212.6267 [astro-ph.CO].
  
\bibitem{Auger:2012ar}
  M.~Auger {\it et al.}  [EXO Collaboration],
  Phys.\ Rev.\ Lett.\  {\bf 109} (2012) 032505
  [arXiv:1205.5608 [hep-ex]].

\bibitem{Blum:2007jz}
  A.~Blum, C.~Hagedorn and M.~Lindner,
  Phys.\ Rev.\ D {\bf 77} (2008) 076004
  [arXiv:0709.3450 [hep-ph]].

\bibitem{Grimus:2004az}
  W.~Grimus and L.~Lavoura,
  J.\ Phys.\ G {\bf 31} (2005) 693
  [hep-ph/0412283].


\bibitem{Zee:1980ai} 
  A.~Zee,
  Phys.\ Lett.\ B {\bf 93}, 389 (1980)
  [Erratum-ibid.\ B {\bf 95}, 461 (1980)].
\bibitem{Wolfenstein:1980sy} 
  L.~Wolfenstein,
  Nucl.\ Phys.\ B {\bf 175}, 93 (1980).
  


\bibitem{Jarlskog:1998uf} 
  C.~Jarlskog, M.~Matsuda, S.~Skadhauge and M.~Tanimoto,
  Phys.\ Lett.\ B {\bf 449}, 240 (1999)
  [hep-ph/9812282].

\bibitem{Chen:2005jm} 
  S.~-L.~Chen, M.~Frigerio and E.~Ma,
  Nucl.\ Phys.\ B {\bf 724}, 423 (2005)
  [hep-ph/0504181].

\bibitem{Frampton:1999yn} 
  P.~H.~Frampton and S.~L.~Glashow,
  Phys.\ Lett.\ B {\bf 461}, 95 (1999)
  [hep-ph/9906375].

\bibitem{He:2003ih} 
  X.~-G.~He,
  Eur.\ Phys.\ J.\ C {\bf 34}, 371 (2004)
  [hep-ph/0307172].


  
\bibitem{Frigerio:2004jg}
  M.~Frigerio, S.~Kaneko, E.~Ma and M.~Tanimoto,
  Phys.\ Rev.\ D {\bf 71} (2005) 011901
  [hep-ph/0409187].

\bibitem{Altarelli:2005yx}
  G.~Altarelli and F.~Feruglio,
  Nucl.\ Phys.\ B {\bf 741} (2006) 215
  [hep-ph/0512103].
  X.~-G.~He, Y.~-Y.~Keum and R.~R.~Volkas,
  JHEP {\bf 0604} (2006) 039
  [hep-ph/0601001].
  F.~Feruglio, C.~Hagedorn, Y.~Lin and L.~Merlo,
  Nucl.\ Phys.\ B {\bf 775} (2007) 120
   [Erratum-ibid.\  {\bf 836} (2010) 127]
  [hep-ph/0702194].
  C.~S.~Lam,
  Phys.\ Rev.\ Lett.\  {\bf 101} (2008) 121602
  [arXiv:0804.2622 [hep-ph]].
  F.~Bazzocchi and S.~Morisi,
  Phys.\ Rev.\ D {\bf 80} (2009) 096005
  [arXiv:0811.0345 [hep-ph]].
  W.~Grimus, L.~Lavoura and P.~O.~Ludl,
  J.\ Phys.\ G {\bf 36} (2009) 115007
  [arXiv:0906.2689 [hep-ph]].


\bibitem{Frampton:2002yf}
  P.~H.~Frampton, S.~L.~Glashow and D.~Marfatia,
  Phys.\ Lett.\ B {\bf 536} (2002) 79
  [hep-ph/0201008].
  
\bibitem{Fritzsch:2011qv} 
  H.~Fritzsch, Z.~-z.~Xing and S.~Zhou,
  JHEP {\bf 1109}, 083 (2011)
  [arXiv:1108.4534 [hep-ph]].


\end{thebibliography}
\end{document}